\documentclass[preprint]{aastex62}

\usepackage{xspace}

\newcommand\LAT{\textsl{Fermi}~LAT\xspace}
\newcommand\Fermi{\textsl{Fermi}\xspace}

\def \Msun {M_\odot\xspace}

\usepackage{soul}

\usepackage{amsmath}
\usepackage{rotating}
\graphicspath{{./}{figures/}}

\submitjournal{ApJ}
\shorttitle{Constraints on the Emission of Gamma Rays from M31 with HAWC}
\shortauthors{HAWC Collaboration}
\begin{document}

\title{Constraints on the Emission of Gamma Rays from M31 with HAWC}

\correspondingauthor{Tolga Yapici}
\email{tyapici@ur.rochester.edu}
\correspondingauthor{Chang Dong Rho}
\email{crho2@ur.rochester.edu, cdr397@skku.edu}

\author{A.~Albert}
\affiliation{Physics Division, Los Alamos National Laboratory, Los Alamos, NM, USA }

\author{R.~Alfaro}
\affiliation{Instituto de F{\'i}sica, Universidad Nacional Aut{\'o}noma de M{\'e}xico, Mexico City, Mexico }

\author{C.~Alvarez}
\affiliation{Universidad Aut{\'o}noma de Chiapas, Tuxtla Guti{\'e}rrez, Chiapas, Mexico}

\author{J.C.~Arteaga-Vel{\'a}zquez}
\affiliation{Universidad Michoacana de San Nicol{\'a}s de Hidalgo, Morelia, Mexico }

\author{K.P.~Arunbabu}
\affiliation{Instituto de Geof{\'i}sica, Universidad Nacional Aut{\'o}noma de M{\'e}xico, Mexico City, Mexico }

\author{D.~Avila~Rojas}
\affiliation{Instituto de F{\'i}sica, Universidad Nacional Aut{\'o}noma de M{\'e}xico, Mexico City, Mexico }

\author{H.A.~Ayala~Solares}
\affiliation{Department of Physics, Pennsylvania State University, University Park, PA, USA }

\author{E.~Belmont-Moreno}
\affiliation{Instituto de F{\'i}sica, Universidad Nacional Aut{\'o}noma de M{\'e}xico, Mexico City, Mexico }

\author{S.Y.~BenZvi}
\affiliation{Department of Physics \& Astronomy, University of Rochester, Rochester, NY , USA }

\author{C.~Brisbois}
\affiliation{Department of Physics, Michigan Technological University, Houghton, MI, USA }

\author{K.S.~Caballero-Mora}
\affiliation{Universidad Aut{\'o}noma de Chiapas, Tuxtla Guti{\'e}rrez, Chiapas, Mexico}

\author{T.~Capistr{\'a}n}
\affiliation{Instituto Nacional de Astrof{\'i}sica, {\'O}ptica y Electr{\'o}nica, Puebla, Mexico }

\author{A.~Carrami{\~n}ana}
\affiliation{Instituto Nacional de Astrof{\'i}sica, {\'O}ptica y Electr{\'o}nica, Puebla, Mexico }

\author{S.~Casanova}
\affiliation{Institute of Nuclear Physics Polish Academy of Sciences, PL-31342 IFJ-PAN, Krakow, Poland }

\author{U.~Cotti}
\affiliation{Universidad Michoacana de San Nicol{\'a}s de Hidalgo, Morelia, Mexico }

\author{J.~Cotzomi}
\affiliation{Facultad de Ciencias F{\'i}sico Matem{\'a}ticas, Benem{\'e}rita Universidad Aut{\'o}noma de Puebla, Puebla, Mexico }

\author{S.~Couti{\~n}o~de~Le{\'o}n}
\affiliation{Instituto Nacional de Astrof{\'i}sica, {\'O}ptica y Electr{\'o}nica, Puebla, Mexico }

\author{E.~De~la~Fuente}
\affiliation{Departamento de F{\'i}sica, Centro Universitario de Ciencias Exactase Ingenierias, Universidad de Guadalajara, Guadalajara, Mexico }

\author{C.~de~Le{\'o}n}
\affiliation{Universidad Michoacana de San Nicol{\'a}s de Hidalgo, Morelia, Mexico }

\author{S.~Dichiara}
\affiliation{Instituto de Astronom{\'i}a, Universidad Nacional Aut{\'o}noma de M{\'e}xico, Mexico City, Mexico }

\author{B.L.~Dingus}
\affiliation{Physics Division, Los Alamos National Laboratory, Los Alamos, NM, USA }

\author{M.A.~DuVernois}
\affiliation{Department of Physics, University of Wisconsin-Madison, Madison, WI, USA }

\author{K.~Engel}
\affiliation{Department of Physics, University of Maryland, College Park, MD, USA }

\author{C.~Espinoza}
\affiliation{Instituto de F{\'i}sica, Universidad Nacional Aut{\'o}noma de M{\'e}xico, Mexico City, Mexico }

\author{H.~Fleischhack}
\affiliation{Department of Physics, Michigan Technological University, Houghton, MI, USA }

\author{N.~Fraija}
\affiliation{Instituto de Astronom{\'i}a, Universidad Nacional Aut{\'o}noma de M{\'e}xico, Mexico City, Mexico }

\author{A.~Galv{\'a}n-G{\'a}mez}
\affiliation{Instituto de Astronom{\'i}a, Universidad Nacional Aut{\'o}noma de M{\'e}xico, Mexico City, Mexico }

\author{D.~Garc{\'i}a-Aguilar}
\affiliation{Instituto de F{\'i}sica, Universidad Nacional Aut{\'o}noma de M{\'e}xico, Mexico City, Mexico }

\author{J.A.~Garc{\'i}a-Gonz{\'a}lez}
\affiliation{Instituto de F{\'i}sica, Universidad Nacional Aut{\'o}noma de M{\'e}xico, Mexico City, Mexico }

\author{F.~Garfias}
\affiliation{Instituto de Astronom{\'i}a, Universidad Nacional Aut{\'o}noma de M{\'e}xico, Mexico City, Mexico }

\author{M.M.~Gonz{\'a}lez}
\affiliation{Instituto de Astronom{\'i}a, Universidad Nacional Aut{\'o}noma de M{\'e}xico, Mexico City, Mexico }

\author{J.A.~Goodman}
\affiliation{Department of Physics, University of Maryland, College Park, MD, USA }

\author{J.P.~Harding}
\affiliation{Physics Division, Los Alamos National Laboratory, Los Alamos, NM, USA }

\author{S.~Hernandez}
\affiliation{Instituto de F{\'i}sica, Universidad Nacional Aut{\'o}noma de M{\'e}xico, Mexico City, Mexico }

\author{B.~Hona}
\affiliation{Department of Physics, Michigan Technological University, Houghton, MI, USA }

\author{D.~Huang}
\affiliation{Department of Physics, Michigan Technological University, Houghton, MI, USA }

\author{F.~Hueyotl-Zahuantitla}
\affiliation{Universidad Aut{\'o}noma de Chiapas, Tuxtla Guti{\'e}rrez, Chiapas, Mexico}

\author{P.~H{\"u}ntemeyer}
\affiliation{Department of Physics, Michigan Technological University, Houghton, MI, USA }

\author{A.~Iriarte}
\affiliation{Instituto de Astronom{\'i}a, Universidad Nacional Aut{\'o}noma de M{\'e}xico, Mexico City, Mexico }

\author{A.~Jardin-Blicq}
\affiliation{Max-Planck Institute for Nuclear Physics, 69117 Heidelberg, Germany}

\author{V.~Joshi}
\affiliation{Max-Planck Institute for Nuclear Physics, 69117 Heidelberg, Germany}

\author{D.~Kieda}
\affiliation{Department of Physics and Astronomy, University of Utah, Salt Lake City, UT, USA }

\author{W.H.~Lee}
\affiliation{Instituto de Astronom{\'i}a, Universidad Nacional Aut{\'o}noma de M{\'e}xico, Mexico City, Mexico }

\author{H.~Le{\'o}n~Vargas}
\affiliation{Instituto de F{\'i}sica, Universidad Nacional Aut{\'o}noma de M{\'e}xico, Mexico City, Mexico }

\author{A.L.~Longinotti}
\affiliation{Instituto Nacional de Astrof{\'i}sica, {\'O}ptica y Electr{\'o}nica, Puebla, Mexico }

\author{G.~Luis-Raya}
\affiliation{Universidad Politecnica de Pachuca, Pachuca, Hgo, Mexico }

\author{K.~Malone}
\affiliation{Physics Division, Los Alamos National Laboratory, Los Alamos, NM, USA }

\author{S.S.~Marinelli}
\affiliation{Department of Physics and Astronomy, Michigan State University, East Lansing, MI, USA }

\author{O.~Martinez}
\affiliation{Facultad de Ciencias F{\'i}sico Matem{\'a}ticas, Benem{\'e}rita Universidad Aut{\'o}noma de Puebla, Puebla, Mexico }

\author{I.~Martinez-Castellanos}
\affiliation{Department of Physics, University of Maryland, College Park, MD, USA }

\author{J.~Mart{\'i}nez-Castro}
\affiliation{Centro de Investigaci{\'o}n en Computaci{\'o}n, Instituto Polit{\'e}cnico Nacional, Mexico City, Mexico.}

\author{J.A.~Matthews}
\affiliation{Dept of Physics and Astronomy, University of New Mexico, Albuquerque, NM, USA }

\author{P.~Miranda-Romagnoli}
\affiliation{Universidad Aut{\'o}noma del Estado de Hidalgo, Pachuca, Mexico }

\author{J.A.~Morales-Soto}
\affiliation{Universidad Michoacana de San Nicol{\'a}s de Hidalgo, Morelia, Mexico }

\author{E.~Moreno}
\affiliation{Facultad de Ciencias F{\'i}sico Matem{\'a}ticas, Benem{\'e}rita Universidad Aut{\'o}noma de Puebla, Puebla, Mexico }

\author{M.~Mostaf{\'a}}
\affiliation{Department of Physics, Pennsylvania State University, University Park, PA, USA }

\author{A.~Nayerhoda}
\affiliation{Institute of Nuclear Physics Polish Academy of Sciences, PL-31342 IFJ-PAN, Krakow, Poland }

\author{L.~Nellen}
\affiliation{Instituto de Ciencias Nucleares, Universidad Nacional Aut{\'o}noma de Mexico, Mexico City, Mexico }

\author{M.~Newbold}
\affiliation{Department of Physics and Astronomy, University of Utah, Salt Lake City, UT, USA }

\author{R.~Noriega-Papaqui}
\affiliation{Universidad Aut{\'o}noma del Estado de Hidalgo, Pachuca, Mexico }

\author{A.~Peisker}
\affiliation{Department of Physics and Astronomy, Michigan State University, East Lansing, MI, USA }

\author{E.G.~P{\'e}rez-P{\'e}rez}
\affiliation{Universidad Politecnica de Pachuca, Pachuca, Hgo, Mexico }

\author{Z.~Ren}
\affiliation{Dept of Physics and Astronomy, University of New Mexico, Albuquerque, NM, USA }

\author{C.D.~Rho}
\altaffiliation{Now at Natural Science Research Institute, University of Seoul, Seoul, Republic of Korea}
\affiliation{Department of Physics \& Astronomy, University of Rochester, Rochester, NY , USA }

\author{D.~Rosa-Gonz{\'a}lez}
\affiliation{Instituto Nacional de Astrof{\'i}sica, {\'O}ptica y Electr{\'o}nica, Puebla, Mexico }

\author{M.~Rosenberg}
\affiliation{Department of Physics, Pennsylvania State University, University Park, PA, USA }

\author{R.~Rubenzahl}
\affiliation{Department of Astronomy, California Institute of Technology, Pasadena, CA, USA }

\author{H.~Salazar}
\affiliation{Facultad de Ciencias F{\'i}sico Matem{\'a}ticas, Benem{\'e}rita Universidad Aut{\'o}noma de Puebla, Puebla, Mexico }

\author{F.~Salesa~Greus}
\affiliation{Institute of Nuclear Physics Polish Academy of Sciences, PL-31342 IFJ-PAN, Krakow, Poland }

\author{A.~Sandoval}
\affiliation{Instituto de F{\'i}sica, Universidad Nacional Aut{\'o}noma de M{\'e}xico, Mexico City, Mexico }

\author{M.~Schneider}
\affiliation{Department of Physics, University of Maryland, College Park, MD, USA }

\author{G.~Sinnis}
\affiliation{Physics Division, Los Alamos National Laboratory, Los Alamos, NM, USA }

\author{A.J.~Smith}
\affiliation{Department of Physics, University of Maryland, College Park, MD, USA }

\author{R.W.~Springer}
\affiliation{Department of Physics and Astronomy, University of Utah, Salt Lake City, UT, USA }

\author{P.~Surajbali}
\affiliation{Max-Planck Institute for Nuclear Physics, 69117 Heidelberg, Germany}

\author{E.~Tabachnick}
\affiliation{Department of Physics, University of Maryland, College Park, MD, USA }

\author{O.~Tibolla}
\affiliation{Universidad Politecnica de Pachuca, Pachuca, Hgo, Mexico }

\author{K.~Tollefson}
\affiliation{Department of Physics and Astronomy, Michigan State University, East Lansing, MI, USA }

\author{I.~Torres}
\affiliation{Instituto Nacional de Astrof{\'i}sica, {\'O}ptica y Electr{\'o}nica, Puebla, Mexico }

\author{L.~Villase{\~n}or}
\affiliation{Facultad de Ciencias F{\'i}sico Matem{\'a}ticas, Benem{\'e}rita Universidad Aut{\'o}noma de Puebla, Puebla, Mexico }

\author{J.~Wood}
\affiliation{Department of Physics, University of Wisconsin-Madison, Madison, WI, USA }

\author{T.~Yapici}
\affiliation{Department of Physics \& Astronomy, University of Rochester, Rochester, NY , USA }

\author{A.~Zepeda}
\affiliation{Physics Department, Centro de Investigacion y de Estudios Avanzados del IPN, Mexico City, Mexico }

\author{H.~Zhou}
\affiliation{Physics Division, Los Alamos National Laboratory, Los Alamos, NM, USA }

%% Mark off the abstract in the ``abstract'' environment. 
\begin{abstract}
Cosmic rays, along with stellar radiation and magnetic fields, are known to make up a significant fraction of the energy density of galaxies such as the Milky Way. When cosmic rays interact in the interstellar medium, they produce gamma-ray emission which provides an important indication of how the cosmic rays propagate.  Gamma rays from the Andromeda Galaxy (M31), located 785~kpc away, provide a unique opportunity to study cosmic-ray acceleration and diffusion in a galaxy with a structure and evolution very similar to the Milky Way.  Using 33 months of data from the High Altitude Water Cherenkov Observatory, we search for TeV gamma rays from the galactic plane of M31. We also investigate past and present evidence of galactic activity in M31 by searching for \Fermi Bubble-like structures above and below the galactic nucleus. No significant gamma-ray emission is observed, so we use the null result to compute upper limits on the energy density of cosmic rays $>10$~TeV in M31.
\end{abstract}


\section{Introduction} \label{sec:intro}

The study of the properties of the Milky Way is unique in that it provides the opportunity to observe the features of the galaxy from within, but at the same time it is challenging to obtain a global perspective of Galactic properties.  Messier 31 (M31), also called the Andromeda galaxy, is the closest neighboring major galaxy to the Milky Way and it is known to be a spiral galaxy of similar structure to the Milky Way. Moreover, M31 is one of the seven external star-forming galaxies that have been observed in GeV gamma rays \citep{Abdo:2010kib,Ackermann:2017nya}. M31 also differs from the Milky Way in important ways: for example, its total stellar mass is a factor of $1.1$ to $2$ larger than our own galaxy \citep{Yin:2009, Licquia:2014rsa, Sick:2014iau}, while its star formation rate is a factor of 2 to 5 lower \citep{Yin:2009, Azimlu:2011aj, Licquia:2014rsa}, possibly decreasing from a peak in the past 10 to 100 Myr \citep{Kang:2009apj}. The similarities and differences make M31 an excellent complement for the study of astrophysics in the Milky Way.

In galaxies such as the Milky Way and M31, cosmic rays can have energy densities comparable to the contributions from magnetic fields and radiation \citep{Garmire:1965}. Thus, they are likely to have a significant influence on galactic dynamics. Moreover, cosmic rays are potential probes of other poorly understood galactic properties such as the star formation rate (SFR).  Due to the presence of Galactic and extragalactic magnetic fields, it is not possible to directly associate charged cosmic rays measured at Earth with acceleration sites in the Milky Way or in other galaxies such as M31. However, cosmic ray interactions with radiation and dense material near their production sites can produce high energy photons, allowing us to probe particle acceleration indirectly. Charged particle acceleration can easily produce spectrally ``hard'' gamma-ray emission extending to TeV energies \citep{Dwek:2012nb}. The gamma-ray flux from a galaxy provides an indirect measurement of the energy density of cosmic rays in the galaxy. With sufficient data, this measurement can then be used to estimate the star formation rate.

For example, the \Fermi~Large Area Telescope (\LAT) has detected a $\sim 10\sigma$ excess of GeV gamma rays with a spectral index of $\Gamma$ of -2.4$\pm$0.1 from the core of M31, and they found agreement between the observed gamma-ray energy flux \citep{Ackermann:2017nya} and the calculated SFR of 0.35 to 1.0 $\Msun$~yr$^{-1}$ by \cite{Yin:2009}. M31 has also been studied for its cosmic ray energy density and other properties by \cite{Yoast-Hull:2015bfk}.  Recently, the Milky Way was found to have undergone past periods of galactic activity via the discovery of the ``\Fermi Bubbles'' \citep{Su:2010qj, Fermi-LAT:2014sfa}. The \Fermi Bubbles (FBs) were first detected during a search for residual gamma rays between 1 and 100~GeV measured by the \LAT \citep{Su:2010qj}. The bubbles subtend approximately 0.8~sr above and below the Galactic Center and are located 9.4~kpc from the Sun under the assumption that they originate from the Galactic Center \citep{Fermi-LAT:2014sfa}. Since most spiral galaxies are believed to contain supermassive black holes at their centers, they may also undergo periods of high activity that give rise to bubble-like structures. Therefore, in the very high energy (VHE) gamma-ray analysis of M31 presented in this paper, we take into consideration the possibility that M31 has its own set of FBs.  We assume a simple morphological model in our analysis based on a study of data from the \LAT by \cite{Pshirkov:2016qhu}, which indicated that M31 does contain two $6-7.5$~kpc circular regions of gamma-ray emission between $0.1-300$~GeV. We note that this emission was not reported in subsequent publication by the \LAT collaboration \citep{Ackermann:2017nya}.  

At TeV energies, the VERITAS Collaboration collected data from several regions around M31 for 54 hours, searching for evidence of VHE emission from the M31 galactic plane and the bubble regions \citep{Bird:2015npa}. However, no significant emission was observed. VERITAS has excellent energy and angular resolution and is very sensitive to point sources, but M31 is a spatially extended source covering roughly $3.2^\circ \times 1^\circ$ \citep{Bird:2015npa}, making observations with VERITAS difficult. On the other hand, the High Altitude Water Cherenkov (HAWC) observatory is well suited to perform unbiased measurements of gamma rays from spatially extended objects.  HAWC is conducting a survey of 2/3 of the sky in gamma rays and is sensitive to gamma rays with energies ranging from 1 TeV to a few hundred TeV \citep{Abeysekara:2017mjj,Abeysekara:2017hyn,Abeysekara:2019edl}. HAWC is located at 4100 m above sea level, at a latitude of 18$^\circ$59.7' N and longitude of 97$^\circ$18.6' W. The nucleus of M31 is located at
$\alpha_{J2000} = 0^h~43^m~35^s.43, \delta_{J2000} = +41^\circ~20'~56.8''$,
and the galaxy transits $22^\circ$ from the zenith of the HAWC detector.

In this paper, we present a systematic search of the extended gamma-ray emission from M31 disk and its possible bubble-like regions above 1~TeV using 33 months of data from HAWC. The paper is structured as follows: we present the spatial models of M31, its possible associated bubbles, and analysis method in Section \ref{sec:method}; in Section \ref{sec:results} we provide gamma-ray upper limits for M31, the bubbles and the combined morphology. We also constrain the cosmic-ray energy density in the M31 disk in Section \ref{sec:results}.

\section{Method}
\label{sec:method}

The analysis of gamma ray sources with HAWC is described in detail in
\citet{Abeysekara:2017mjj}. In brief, the energy of each gamma ray is estimated
from the fraction of the optical sensors in HAWC triggered by gamma-ray air
showers. To estimate the energy spectrum of an observed source of gamma rays,
parametric spatial and spectral models of gamma ray sources are fit to HAWC
data using a maximum likelihood technique. We employ the Multi-Mission Maximum
Likelihood analysis framework \citep{Vianello:2015wwa, Younk:2015saa} to fit parametric spatial and spectral models to the HAWC data. For both the M31 disk and bubble regions, we assume the gamma-ray emission is described by a simple power law
\begin{equation}
\frac{dN}{dE} = K\left(\frac{E}{E_0}\right)^{\Gamma},
\label{eq:powerlaw}
\end{equation}
where $K$ is a normalization constant, $E_0$ is the pivot energy, and $\Gamma$ is the spectral index. For the disk-only analysis, we perform a maximum likelihood fit of the flux normalization with three different fixed spectral indices. The same procedure is followed for the bubbles-only analysis. For the combined case, we perform the analysis for all combinations of spectral indices for the disk and the bubble regions. Note that when fitting, the normalizations and the spectral indices are assumed to be same for both bubbles. Throughout the analysis, a fixed pivot energy of $E_0=1$~TeV is used and the spectral indices of $\left[-2.0,\, -2.5,\, -3.0\right]$ are considered. Our estimate of statistical significance, the ``Test Statistic'' (TS), is computed for each fit as follows \citep{Albert:2017vtb, Wilks:1938}:

\begin{eqnarray}
    \rm{TS} & = & 2\ln\frac{\mathcal{L}(\text{source(s)}+\text{background})}{\mathcal{L}(\text{background})} \\
    \text{significance} & \approx & \sqrt{\rm{TS}}~[\text{Gaussian $\sigma$}]
    \label{eqn:TS}
\end{eqnarray}

The maximum-likelihood estimators are used to find the best normalizations,
$\hat{K}_\text{disk}$ and $\hat{K}_\text{FB}$. We then use
$\hat{K}_\text{disk}$ and $\hat{K}_\text{FB}$ as input to a Markov-Chain
Monte Carlo (MCMC) in order to estimate the distribution of the posterior
likelihood around the maximum. In the MCMC, a uniform prior range is assumed for
both $K_\text{disk}$ and $K_\text{FB}$ between $0$ and
$10^{-11}$~TeV$^{-1}$~cm$^{-2}$~s$^{-1}$. Given the lack of statistically
significant VHE emission, the MCMC results are then used to compute 95\%
credible upper intervals on the flux normalization.

In addition to upper limits on $K$ assuming a fixed $\Gamma$ for the disk,
bubbles, and disk and bubbles in combination, we also compute
quasi-differential limits between $1$ and $100$~TeV. That is, the data are
sorted into five energy bins $j\in[1,\ldots,5]$ and independent power-law fits
are used to constrain $K_j$ and $\Gamma_j$ in each bin. The bin widths
correspond to a half-decade in gamma-ray energy, following the analysis of HAWC
data presented in \citet{Aartsen:2017snx}, with the pivot energy $E_{0,j}$ is set to the center of the logarithmic energy range of bin $j$. The energy bins are defined in Table~\ref{table:ebins}.

\begin{table}[h!]
    \centering
    \begin{tabular}{c c c c}
      Energy bin & Lower Energy [TeV] & Higher Energy [TeV] & Pivot Energy ($E_0$) [TeV] \\
        \hline \hline
        1          & 0.56         & 1.78          & 1.0           \\
        2          & 1.78         & 5.62          & 3.16          \\
        3          & 5.62         & 17.78         & 10.0          \\
        4          & 17.78        & 56.23         & 31.63         \\
        5          & 56.23        & 177.8         & 100.0         \\
        \hline
    \end{tabular}
    \caption{Definition of energy bins used for the quasi-differential limit calculations, following the analysis used in \cite{Aartsen:2017snx}.}
    \label{table:ebins}
\end{table}
\subsection{Model Templates}
We treat the morphology of the M31 FBs as described in \cite{Pshirkov:2016qhu}. The position of the core of M31 is taken from the 100~$\mu$m infrared map found in the Improved Reprocessing of the IRAS Survey (IRIS) database \citep{MivilleDeschenes:2004ci}. The location of the M31 galactic center, the positions of the bubbles, and the orientation of the disk of M31 used in the analysis are given in Table~\ref{tab:positions}.

\begin{figure}[h!]
\centering
\includegraphics[width=0.45\textwidth]{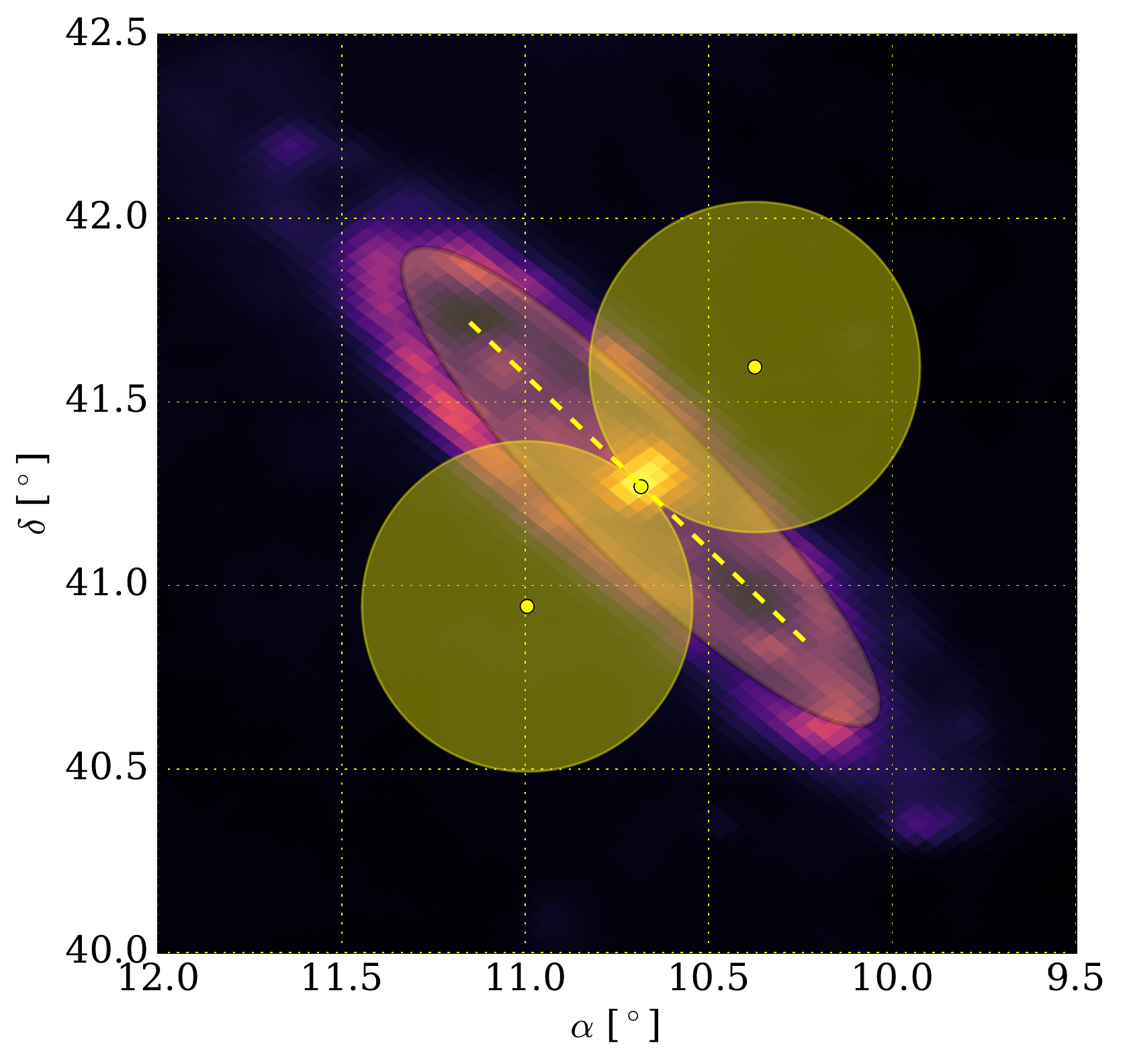}
\caption{Regions to be used in the model for the FBs around M31. The M31 galactic plane is modeled as an ellipse with the two FBs as circular regions with radius $0.45^\circ$. The background image is the IRIS 100$~\mu m$ map of the $2.5^\circ \times 2.5^\circ$ region surrounding M31. The dashed line represents the M31 galactic plane between the two endpoints defined in Table~\ref{tab:positions}.}
\label{fig:bubblemodel}
\end{figure}

The M31 galactic disk is modeled as an ellipse with a semi-major axis of length $0.9^\circ$ inclined 45.04$^{\circ}$ counterclockwise from the positive right ascension axis. The ellipse has an aspect ratio of $b/a = 0.22$ \citep{Pshirkov:2016qhu}. The two FB regions are modeled as circular disks of radius $0.45^\circ$, oriented perpendicular to and symmetrically above and below the galactic plane of M31. Figure~\ref{fig:bubblemodel} displays the regions used in the spatial fit. Two of the model templates used in the analysis are based on M31 and the FB separately. The third model template is the combination of M31 and the FBs. The flux is assumed to be constant within the area encompassed by the template.

\begin{table}[h!]
\centering
\begin{tabular*}{0.75\textwidth}{@{\extracolsep{\fill}} c c c }
\hspace{10pt}Region & $\alpha_{J2000}$ & $\delta_{J2000}$\hspace{10pt} \\
\hline\hline% \hline
\hspace{10pt}M31 Nucleus & $10.6848^\circ$ & $41.7166^\circ$\hspace{10pt} \\
%\hline
\hspace{10pt}M31 Disk Endpoint 1 & $11.1509^\circ$ & $40.8256^\circ$\hspace{10pt} \\  
%\hline
\hspace{10pt}M31 Disk Endpoint 2 & $10.2145^\circ$ & $40.8256^\circ$\hspace{10pt} \\
%\hline
\hspace{10pt}FB 1 Center & $10.3746^\circ$ & $41.5951^\circ$\hspace{10pt} \\
%\hline
\hspace{10pt}FB 2 Center & $10.9949^\circ$ & $40.9430^\circ$\hspace{10pt} \\
\hline
\end{tabular*}
\caption{Positions of M31 regions from the IRIS database \citep{MivilleDeschenes:2004ci,Pshirkov:2016qhu}. The two endpoints and nucleus of M31 define the galactic plane, which we use to locate two positions perpendicular and symmetric to the center of the plane (FB1 and FB2) that are spaced $0.45^\circ$ from the nucleus.}
\label{tab:positions}
\end{table}

\subsection{Cosmic-ray Energy Density}\label{subsec:cr_density}

Using a measurement of or constraint on the flux of TeV gamma rays, under the
assumption that gamma-ray emission is due to the decay of neutral pions
produced by cosmic-ray interactions, the energy density of cosmic rays
responsible for the production of gamma rays can be derived using
\citep{Abramowski:2016mir}
\begin{eqnarray}
\omega_{CR}\left(\geq 10E_{\gamma}\right) & \sim &
1.8\times10^{-2}\left(\frac{\eta_{n}}{1.5}\right)^{-1}\left(\frac{L_{\gamma}\left(\geq
E_{\gamma}\right)}{10^{34}~\text{erg~s$^{-1}$}}\right)\left(\frac{M}{10^{6}~\Msun}\right)^{-1}~\text{eV~cm$^{-3}$}.
\label{eqn:cosmic}
\end{eqnarray}
In eq.~\ref{eqn:cosmic}, $L_{\gamma}\left(\geq E_{\gamma}\right)$ is the
observed gamma-ray luminosity at energies beyond $E_\gamma$, and $M$ is the
mass of target nuclei within the region of interest.  Note that $\omega_{CR}$
is estimated for $\geq 10E_{\gamma}$ because, on average, approximately 10\% of
the primary cosmic-ray energy goes into the production of gamma-ray photons via
the interactions $pp\to pp\pi^0$ and $\pi^0\to\gamma\gamma$.  Estimates of target mass are
typically based on measurements of HI and H$_2$; the quantity $\eta_{n}$ is a
scale factor fixed at 1.5 to account for nucleons present other than hydrogen
\citep{Abramowski:2016mir}. Since physical properties such as the mass of the
galaxy are required for the calculation, the cosmic-ray energy density is only
computed for M31 and not for the bubbles. In addition, we also compute the
cosmic-ray energy density corresponding to a gamma-ray energy above 1~TeV for
each energy bin adopted from the quasi-differential limits. 

Current best estimates of the properties of M31 are used with relevant
uncertainties (see Table \ref{table:values_used}). The uncertainties are
propagated to calculate the systematic uncertainties on each value of
$\omega_\text{CR}$. Most of the uncertainties on the energy densities come from
the estimates of the target mass and the detector systematics in the HAWC
analysis \citep{Abeysekara:2017mjj}.

\begin{table}[hb]
\centering
\begin{tabular}{ c c c }
\hline
Mass HI          & {\bf $(7.33\pm2.20)\times10^{9}~\Msun$} & {\bf \cite{Braun:2009apj}} \\
Mass H$_2$          & {\bf $(3.45\pm1.80)\times10^{9}~\Msun$} & {\bf \cite{Nieten:2005if}} \\
Distance         & $d=785\pm25~\text{kpc}$            & \cite{McConnachie:2004dv}   \\
Flux Uncertainty & $\pm$50\%                            & \cite{Abeysekara:2017mjj} \\
\hline
\end{tabular}
\caption{The mass budget used to calculate the cosmic-ray energy distribution.}
\label{table:values_used}
\end{table}

\section{Results}
\label{sec:results}

The analysis uses data recorded with HAWC between June 2015 and December 2017
\citep{Abeysekara:2017mjj}.  A description of the construction of HAWC sky maps
is provided in \cite{Abeysekara:2017hyn}. In brief, maps of the sky around M31
are produced by binning the arrival directions of cosmic rays and gamma rays in
a HEALPix map \citep{Gorski:2004by} using bins of width 0.057$^\circ$,
considerably smaller than the $0.2^\circ$ to $1^\circ$ angular resolution of
HAWC.  Cosmic ray air showers are identified and rejected based on the variance
of the distribution of charge deposited as a function of distance from the
reconstructed shower core. The rejection power of the cosmic-ray cut depends
strongly on the size of the air shower; 10\% of cosmic rays below 1~TeV pass
the event selection, while 0.1\% pass at 10~TeV \citep{Abeysekara:2017mjj}. The
gamma-ray selection efficiency is $>30\%$ at all energies.

\begin{figure}[h!]
  \includegraphics[width=0.325\textwidth]{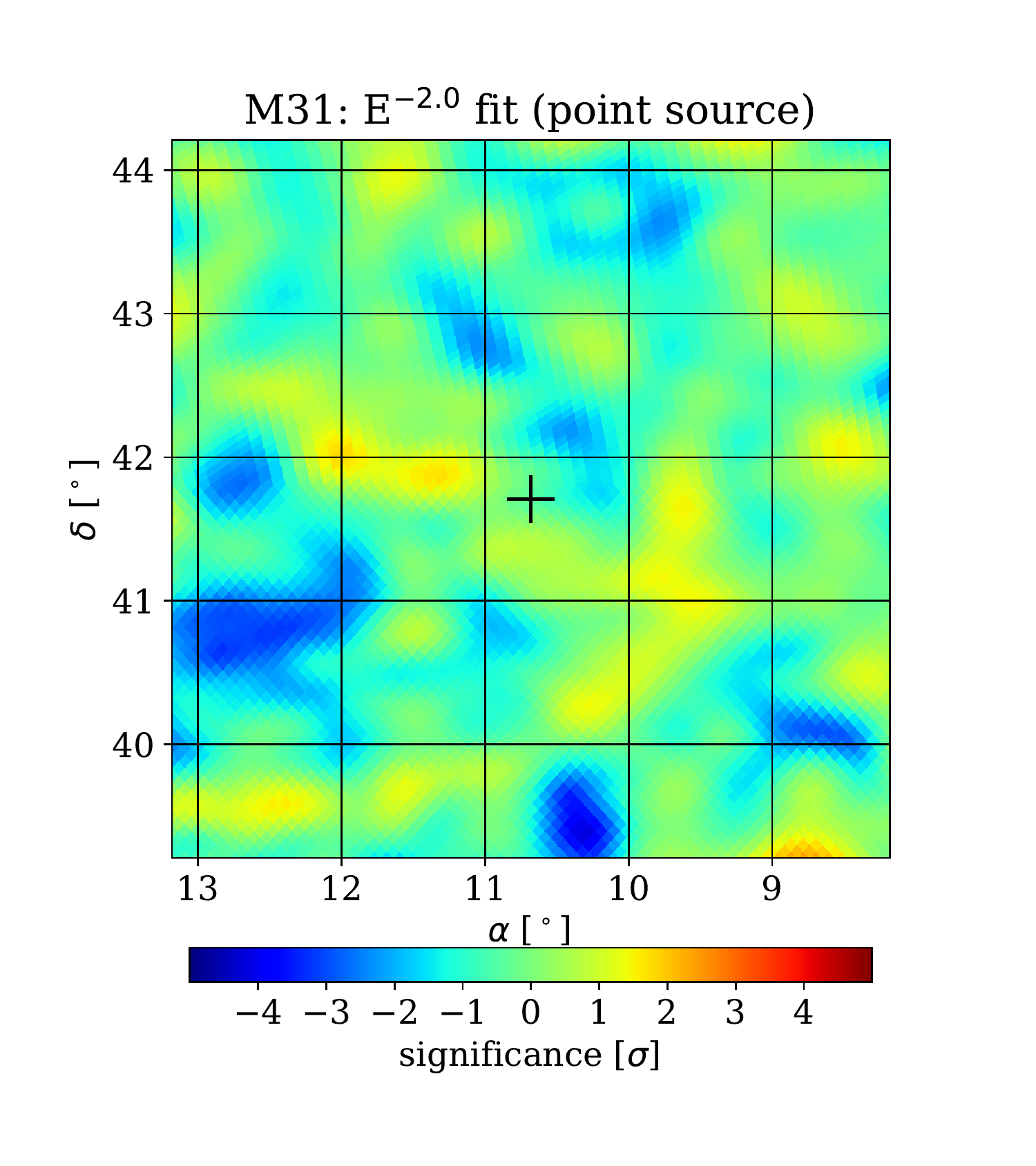}
  \hfill
  \includegraphics[width=0.325\textwidth]{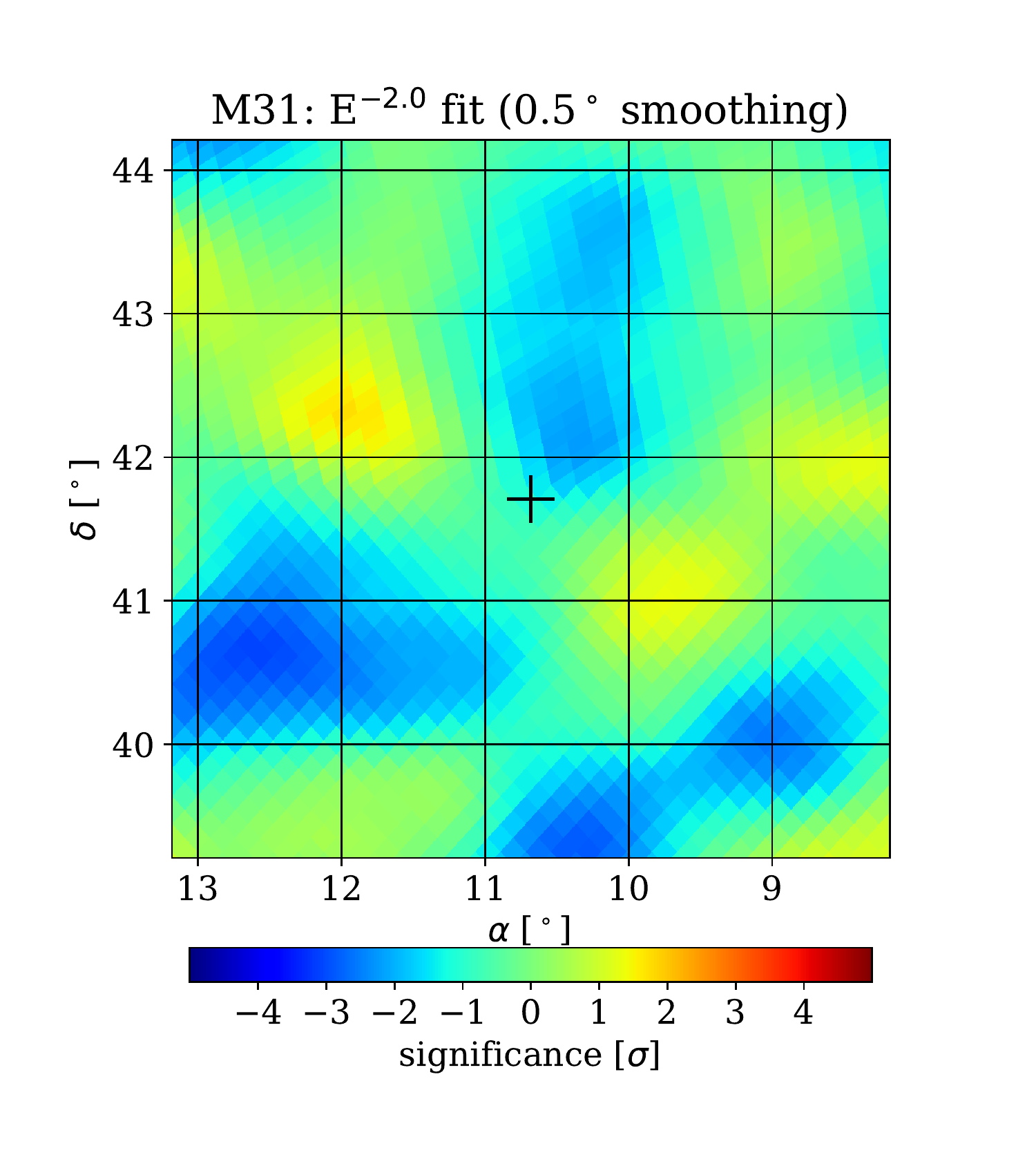}
  \hfill
  \includegraphics[width=0.325\textwidth]{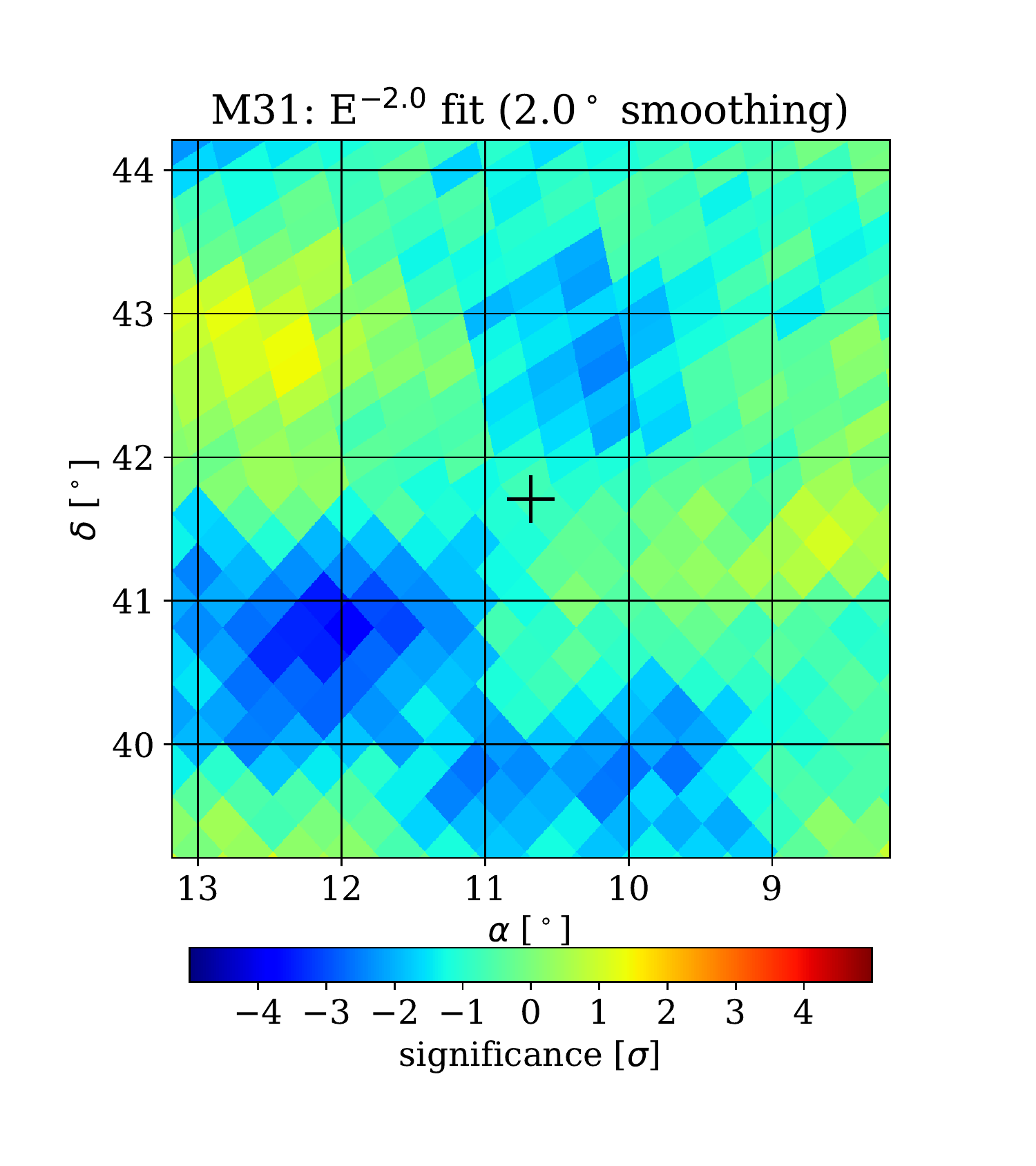}
  \caption{\label{fig:m31_sig_map}
  Significance map around M31 (crosshairs) produced with 33 months of HAWC data using (left) the point source map, (center) $0.5^\circ$-smoothed map and (right) $2^\circ$-smoothed map. An $E^{-2}$ spectrum is used for the three maps.}
\end{figure}

Once the HEALPix map is filled, we produce a background map giving the expected number of counts from an isotropic distribution of events. The background expectation is computed from the data themselves using the direct integration method developed in \cite{Atkins:2003ep}. A maximum likelihood technique is then used to search for significant excesses and deficits with respect to the isotropic background expectation as a function of sky location. The signal model in the likelihood assumes each HEALPix pixel includes background counts plus a flux from a power-law energy spectrum. The source or sources contributing to counts in a given pixel can be treated as point-like or spatially extended, assuming a uniform disk of fixed radius; the source morphology is convolved with the detector angular resolution in the likelihood. 

Figure~\ref{fig:m31_sig_map} shows HAWC sky maps around M31 assuming an $E^{-2}$ source spectrum and a point-like source morphology (left), a source extended by $0.5^\circ$ (center), and a $2^\circ$ extended source (right).  The maps include all energies $>1$~TeV, and no significant excess emission is observed.  Thus, for each of our model templates, we report 95\% credible upper limits on the flux of gamma rays above 1 TeV.

To estimate the sensitivity of our analysis, we calculate the expected upper limits of HAWC to the M31 disk and two FBs (as well as individual components of such a combined model) by fitting the same M31 spatial templates to background-only regions at different right ascensions, keeping the declination of the templates fixed. By avoiding regions with known VHE gamma-ray sources, this produces a distribution of background-only estimates for the particular morphological models we are using.  These are used to provide expected limits of the model templates used in our analysis and 68\% and 95\% containment bands of the expected limits.

\subsection{Expected Gamma-ray Flux from M31 Fermi Bubbles, assuming Milky-Way like Emission}
\label{sec:extrapolation}

\begin{figure}[b!]
\centering
\includegraphics[width=\textwidth]{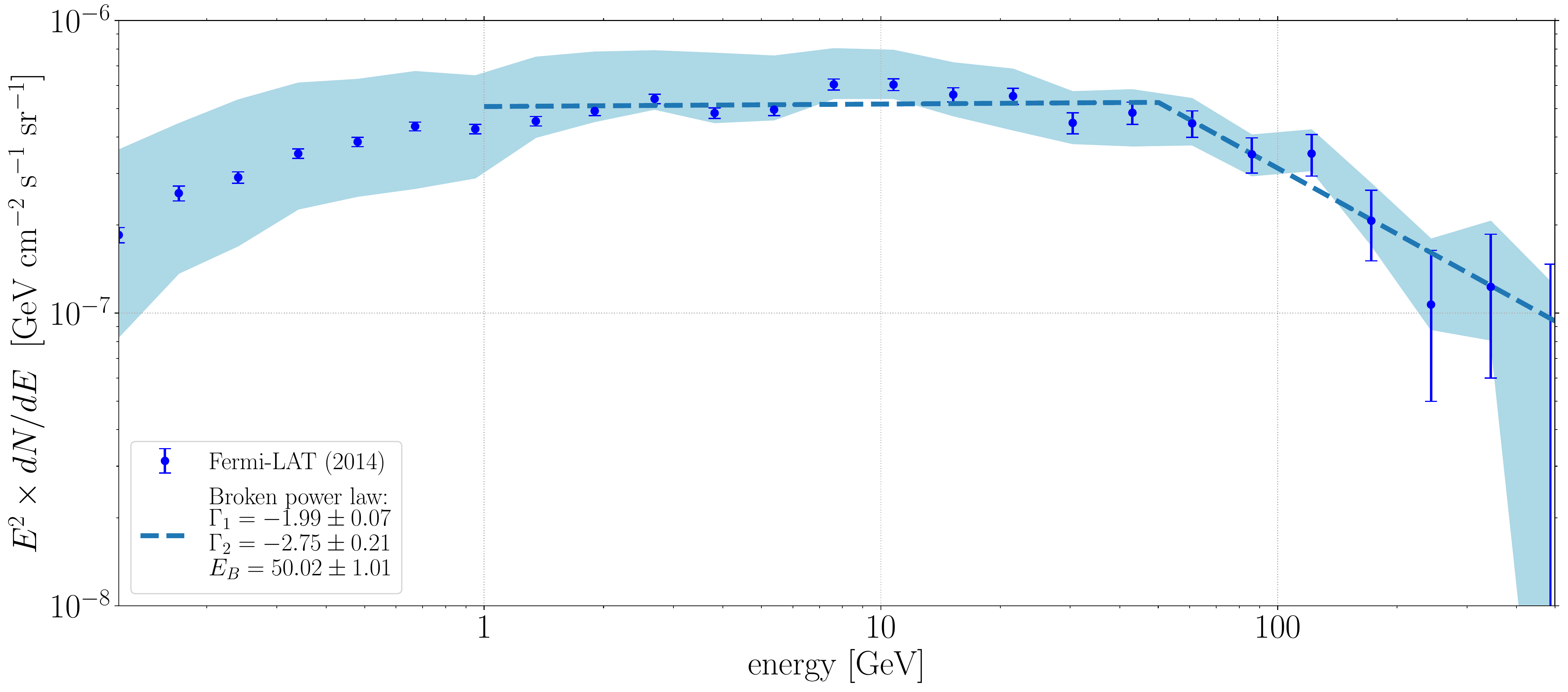}
\caption{Spectral energy distribution (SED) of the Milky Way \Fermi Bubbles,
with statistical uncertainties shown as error bars and systematic uncertainties
as the filled contour \citep{Fermi-LAT:2014sfa}. The dashed line is a best-fit
broken power law between $1-500$ GeV with spectral indices $\Gamma_1$,
$\Gamma_2$, and break energy $E_B$. Below 50~GeV the flux is an $E^{-2}$
spectrum; above the spectrum is described by $E^{-2.75}$.}
\label{fig:SED}
\end{figure}

The observed spectral energy distribution (SED) of the Milky Way FBs reported by \cite{Fermi-LAT:2014sfa} is approximately constant in the 1~-~100~GeV energy range. However, at higher energies the spectrum falls off as shown in Figure~\ref{fig:SED}.

From the observed flux of the Milky Way bubbles, we can calculate the gamma-ray
flux of equivalent bubble-like structures of M31, given the distance
$d_\text{M31}=785$~kpc between M31 and the Milky Way. This follows a similar
analysis carried out by \cite{Pshirkov:2016qhu}. Since HAWC is sensitive to
gamma rays in the TeV regime, we extrapolate the SED to energies greater than
1~TeV. First, we calculate the integral flux of the Milky Way FBs by
integrating the best-fit power law (dashed line in Figure~\ref{fig:SED}) above
1~TeV multiplied by the total solid angle of the bubbles. To extrapolate the
flux to what is expected from a similarly sized source at the position of M31,
we scale the integral flux by ($d_\text{MW}$/$d_\text{M31}$)$^2$ where
$d_\text{MW}$ is the assumed distance to the Milky Way FBs. Under these assumptions, the expected integral flux from M31 FBs is approximately $3.9\times10^{-15}$~cm$^{-2}$~s$^{-1}$ at energies $>1$~TeV.

Using the constant flux from 1 to 50 GeV (the dashed curve in
Fig.~\ref{fig:SED}), we calculate the total integral flux expected from M31
bubbles to be $2.06\times10^{-10}$~cm$^{-2}$~s$^{-1}$. This is significantly
below the integral flux of $(2.6\pm0.6)\times10^{-9}$~cm$^{-2}$~s$^{-1}$ one
obtains using the spectral index $\Gamma_\text{FB}=2.3\pm0.1$ reported by
\cite{Pshirkov:2016qhu}. The \LAT Collaboration does not report any
statistically significant GeV emission from the bubble regions
\citep{Ackermann:2017nya}.

\subsection{Spectral Analysis}

\begin{figure}
    \centering
    \includegraphics[width=\textwidth]{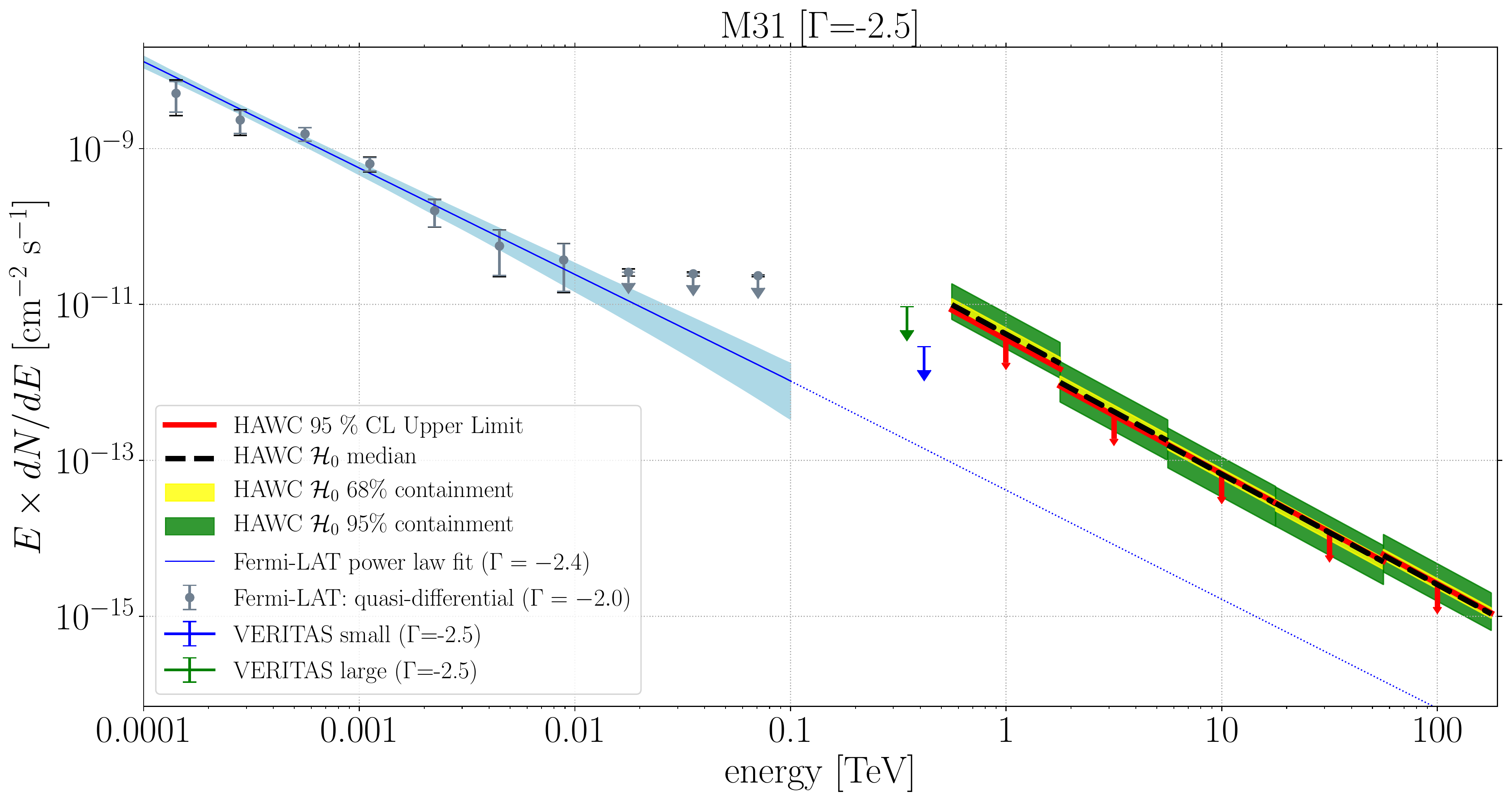}
    \caption{\label{fig:m31_results} 95\% credible upper intervals on
    gamma-ray emission from the disk of M31 using the 33-month dataset from
    HAWC (red arrows), assuming an $E^{-2.5}$ power law. The expected range of
    upper limits assuming the null hypothesis of no emission ($\mathcal{H}_0$)
    was computed using background-only regions.
    For each quasi-differential energy bin, dotted lines represent mean upper
    limits, yellow boxes represent 
    68\% containment regions for the upper limits, and green boxes represent 
    95\% containment regions.
    (See Section~\ref{sec:results} for details.)
    Also shown are measurements of
    M31 from \LAT \citep{Ackermann:2017nya} and upper limits from VERITAS
    \citep{Bird:2015npa}.}
\end{figure}

\begin{figure}
    \centering
    \includegraphics[width=\textwidth]{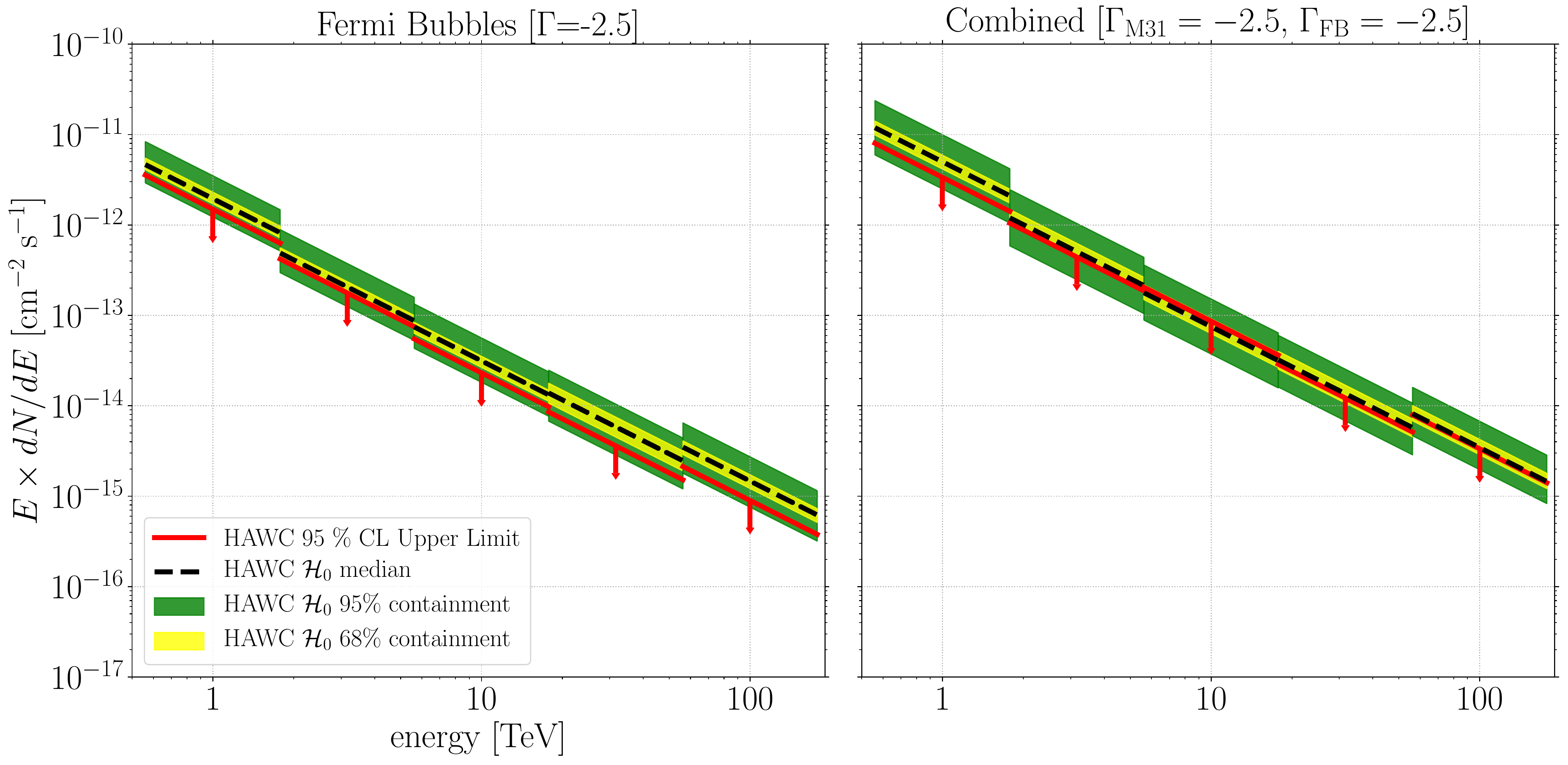}
    \caption{\label{fig:fb_results} {\sl Left}: 95\% credible upper limits on
    VHE emission from ``\Fermi Bubbles'' in M31, assuming an $E^{-2.5}$
    spectrum. {\sl Right}: 95\% upper limits VHE emission from a joint fit of
    M31 and the FB regions, assuming independent spectral normalizations but
    the same power law index of $-2.5$.}
\end{figure}

The spectrum of the M31 disk, the bubbles and the combined spatial model are
shown in Figures \ref{fig:m31_results} and \ref{fig:fb_results}. We summarize
integral flux upper limits for the M31 disk, bubbles and combined models in
Table~\ref{tab:limits} based on power law fits (eq.~\ref{eq:powerlaw}); the
table also includes expected upper limits for the corresponding models.
Figures~\ref{fig:m31_results} and \ref{fig:fb_results} show quasi-differential
limits calculated using the method described in Section~\ref{sec:method}. Plots
based on the single power law fits and tables giving the normalizations $K$ for
each fit are provided in the Appendix of this paper.

For the various single and joint fits, the limits on the VHE integral gamma-ray
flux $>1$~TeV differ by roughly a factor of six when different spectral indices
are assumed. We note that all upper limits are within the 95\% intervals of the
``expected'' upper limits, defined using fits to background-only regions
observed by HAWC (see Section~\ref{sec:method}). Thus the upper limits on the
VHE flux are consistent with the expected sensitivity of the detector. For
comparison, VERITAS upper limits corresponding to a large and a small test
region of M31 \citep{Bird:2015npa} are also shown in
Fig.~\ref{fig:m31_results}..

\begin{table}[h!]
\centering
\begin{tabular*}{\textwidth}{@{\extracolsep{\fill}} c c c }
Model Template  & Integral Flux Upper Limit & Expected Upper Limit \\
& ($>1$ TeV) [$10^{-13}\text{ cm}^{-2}\text{ s}^{-1}$] & ($>1$ TeV) [$10^{-13}\text{ cm}^{-2}\text{ s}^{-1}$] \\
\hline \hline
M31 Disk ($\Gamma = -2.0$) & 0.85 & 0.93 (0.77-1.09, 0.51-1.65) \\
M31 Disk ($\Gamma = -2.5$) & 3.75 & 3.83 (3.18-4.63, 2.08-6.03) \\
M31 Disk ($\Gamma = -3.0$) & 5.18 & 7.10 (6.00-8.77, 4.41-12.85) \\
\hline
FB ($\Gamma = -2.0$) & 0.31 & 0.50 (0.39-0.63, 0.24-0.86) \\
FB ($\Gamma = -2.5$) & 1.67 & 1.91 (1.62-2.27, 1.01-3.31) \\
FB ($\Gamma = -3.0$) & 2.03 & 3.28 (2.86-4.03, 1.96-6.30) \\
\hline
Combined ($\Gamma_{M31} = -2.0, \Gamma_{FB} = -2.0$) & 1.12 & 1.19
(0.97-1.48, 0.53-2.37)   \\
Combined ($\Gamma_{M31} = -2.0, \Gamma_{FB} = -2.5$) & 1.87 & 2.26
(1.78-3.03, 0.95-4.53)  \\
Combined ($\Gamma_{M31} = -2.0, \Gamma_{FB} = -3.0$) & 2.64 & 3.43
(2.63-4.88, 1.26-6.80) \\
Combined ($\Gamma_{M31} = -2.5, \Gamma_{FB} = -2.0$) & 2.97 & 2.98
(2.30-4.17, 1.00-6.71)  \\
Combined ($\Gamma_{M31} = -2.5, \Gamma_{FB} = -2.5$) & 4.20 & 4.59
(3.75-5.84, 2.26-8.89)  \\
Combined ($\Gamma_{M31} = -2.5, \Gamma_{FB} = -3.0$) & 5.54 & 6.08
(4.83-7.40, 2.59-11.39)  \\
Combined ($\Gamma_{M31} = -3.0, \Gamma_{FB} = -2.0$) & 5.25 & 5.36
(4.24-6.57, 2.06-12.55)  \\
Combined ($\Gamma_{M31} = -3.0, \Gamma_{FB} = -2.5$) & 6.04 & 7.17
(5.44-8.77, 3.00-14.41)  \\
Combined ($\Gamma_{M31} = -3.0, \Gamma_{FB} = -3.0$) & 7.97 & 8.36
(6.99-10.19, 4.53-16.42)  \\
\hline
\end{tabular*}
\caption{Observed 95\% credible upper limits on the emission from the M31
galactic disk and its bubbles. The expected median upper limits, as well as the
68\% and 95\% containment bands for the expected limits are given in the last
the column.}
\label{tab:limits}
\end{table}

\subsection{Cosmic Ray Energy Density Interpretation}
The calculated integral flux upper limits are used to infer the measured upper
limits for the cosmic-ray energy density of the M31 disk above 1~TeV. The
results are tabulated in Table~\ref{tab:cr_limits} along with the uncertainties
calculated by propagating the uncertainties in the disk mass and distance (see
Table~\ref{table:values_used}) and the systematic uncertainties on gamma-ray
spectra measured with HAWC \citep{Abeysekara:2017mjj}.

For limits on the quasi-differential cosmic-ray energy density, we calculate
$\omega_{CR}\left(\geq 10E_{\gamma}\right)$ with $E_{\gamma}$ set to the lower
and higher end energy values for each energy bin in measured quasi-differential
flux upper limits. As described in Section~\ref{subsec:cr_density}, $10E_\gamma$ is used as a
proxy for the energy of the hadronic cosmic rays which produce gamma rays via
the production and decay of neutral pions.  Then we use the values to compute
the quasi-differential cosmic-ray energy density upper limits per energy bin
using:

\begin{equation}
    \omega_{CR}\left(10E_{\gamma, 2} \geq 10E_\gamma \geq 10E_{\gamma, 1}\right) = \omega_{CR}\left(\geq 10E_{\gamma, 1}\right) - \omega_{CR}\left(\geq 10E_{\gamma, 2}\right)
\end{equation}

Our limits on the cosmic ray energy density agree with the expected upper
limits computed in background-only regions. For all energy bins considered in
this work, the computed upper limits on $\omega_\mathrm{CR}$ are higher than
the average cosmic ray energy density of the Milky Way calculated by the
H.E.S.S. Collaboration \citep{Abramowski:2016mir}, $\sim10^{-3}$ eV~cm$^{-3}$.

\begin{table}[h!]
\centering
\begin{tabular*}{0.5\textwidth}{@{\extracolsep{\fill}} c c}
               & Upper limit \\
Model Template & $\omega_{CR}$($>$ 10 TeV) \\
               & [eV/cm$^3$]     \\
\hline \hline
%M31 Disk ($\Gamma=-2.0$) & $0.0108 \pm 0.0085$ \\
%M31 Disk ($\Gamma=-2.5$) & $0.0281 \pm 0.0220$ \\
%M31 Disk ($\Gamma=-3.0$) & $0.0284 \pm 0.0223$ \\
M31 Disk ($\Gamma=-2.0$) & $0.011 \pm 0.009$ \\
M31 Disk ($\Gamma=-2.5$) & $0.028 \pm 0.022$ \\
M31 Disk ($\Gamma=-3.0$) & $0.028 \pm 0.022$ \\
\hline
\end{tabular*}
\caption{Calculated upper limits on the cosmic-ray energy density for the M31 disk for three spectral indices.}
\label{tab:cr_limits}
\end{table}

%\begin{figure}[h!]
%    \centering
%    \includegraphics[width=.45\textwidth]{figures/results/cosmic_ray_energy_density_ver2.pdf}
%    \caption{\label{fig:quasi_diff_limits_M31disk} Cosmic-ray energy densities calculated using M31 disk quasi-differential upper limit. Propagated uncertainties are shown with hashed bands.}
%    \label{fig:quasi_cre}
%\end{figure}

\begin{figure}[h!]
    \centering
    \includegraphics[width=.5\textwidth]{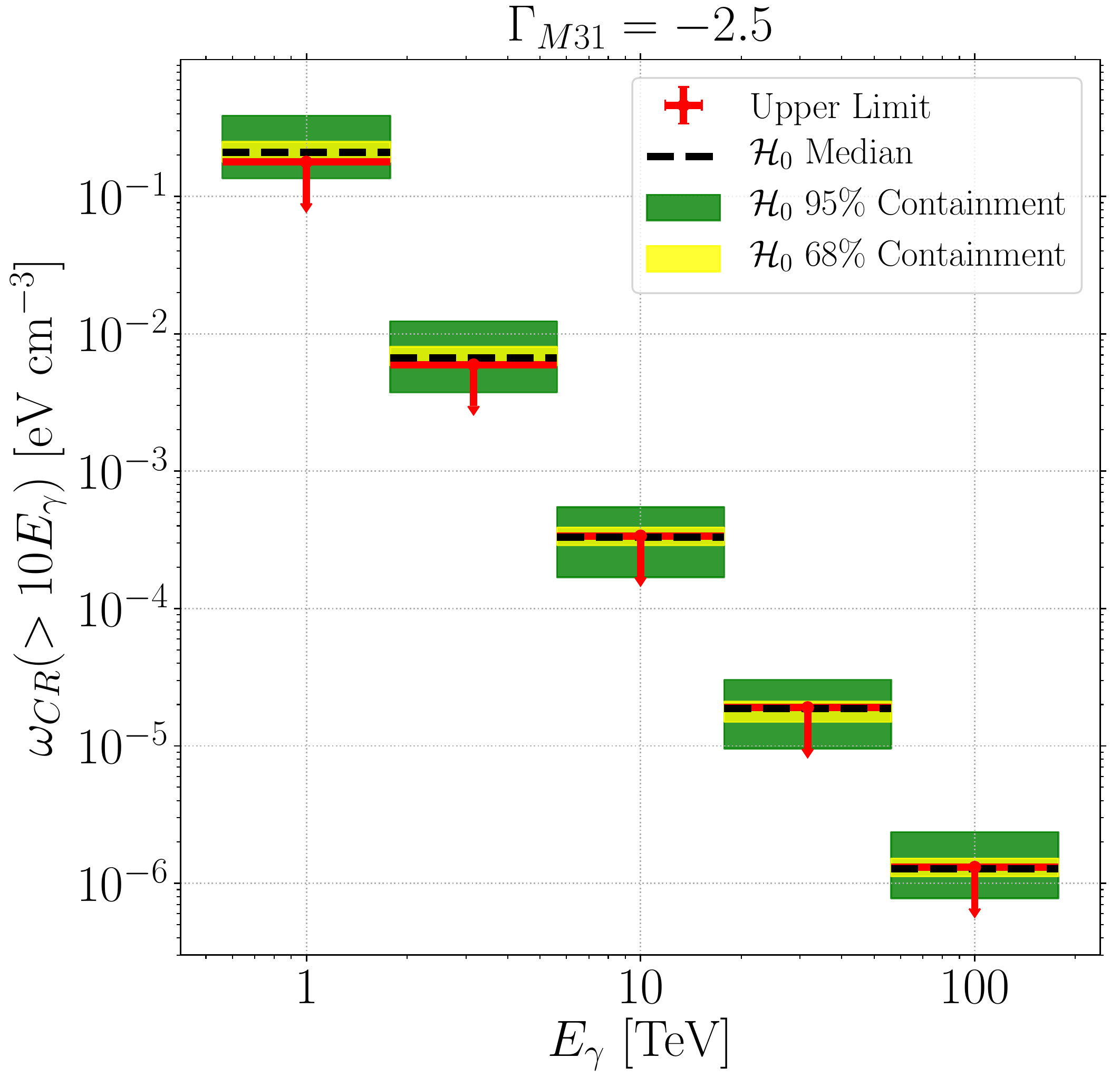}
    \caption{Upper limits on the energy density of cosmic rays in M31 $>10$~TeV
    using the HAWC data set. The calculation assumes an $E^{-2.5}$ spectrum for
    the VHE gamma rays.}% The Milky way cosmic-ray energy density \citep{Abramowski:2016mir} is shown for comparison.}
    \label{fig:cosmicray_sensitivity}
\end{figure}

\section{Conclusions}

We have searched for gamma-ray emission from the nearby spiral galaxy, M31, applying spatial templates with 33 months of HAWC data. Independent and joint fits to the galactic disk and FB-like structures are performed. No significant emission is observed in the regions of M31 with the HAWC point source and extended source searches. Therefore, we report flux upper limits for the disk, the bubbles and the two features combined. The calculated integral flux upper limits are consistent with HAWC's sensitivity to background-only regions. The \LAT spectral analysis of the M31 disk \citep{Ackermann:2017nya} provides a spectral index of $\Gamma$ of -2.4$\pm$0.1. When the spectrum is extrapolated under the assumption of no spectral break, the corresponding particle flux above 1 TeV has a value of 2.51$\times$10$^{-14}$ cm$^{-2}$s$^{-1}$. The flux upper limits from HAWC are about an order of magnitude higher than the extrapolated results, but consistent with the HAWC sensitivity. Our results do not coincide in energy range with VERITAS results \citep{Bird:2015npa}; however, our upper limits have an agreement with their upper limits.

Using the calculated flux upper limits of the M31 galactic disk, we estimate
its cosmic-ray energy density using different fixed spectral indices at
$>$1~TeV extrapolated to the full energy range of HAWC (see Table
\ref{tab:cr_limits}) as well as quasi-differential energy bins (Figure
\ref{fig:cosmicray_sensitivity}). We note that if we extrapolate the M31 disk
spectrum measured by the \LAT
\citep{Ackermann:2017nya} to the $1-100$ TeV range, we obtain an integral flux
of 7.40$\times$10$^{-14}$ TeV cm$^{-2}$~s$^{-1}$. This value
can be interpreted as cosmic ray energy density of 1.05$\times$10$^{-5}$
eV~cm$^{-3}$, which is consistent with the upper limits on cosmic ray energy
density that we estimate with HAWC and report in this work. 

The HAWC limits can also be compared to the flux of the FBs in the Milky Way,
normalized to the distance to M31, as described in
Section~\ref{sec:extrapolation}. The upper limits of the integral flux of the
M31 FB structures (0.30 - 2.68$\times10^{-13}$~cm$^{-2}$s$^{-1}$) are found to
be higher than the extrapolated flux of 3.9$\times10^{-15}$~cm$^{-2}$s$^{-1}$
by approximately two orders of magnitude.

This analysis is the first systematic study of the extended gamma-ray
emission from the M31 region at TeV energies by a detector optimized to study
extended emission. While no VHE gamma rays from the disk of M31 or ``\Fermi
Bubble'' regions around the galactic nucleus were observed, we have constrained
emission from this region up to 100~TeV. Future versions of this analysis will
benefit from a factor of two increase in the size of the HAWC data set, as well
as the development of high-resolution gamma-ray energy estimators in the HAWC
reconstruction of gamma-ray air showers.

\section*{Acknowledgements}

We acknowledge the support from: the US National Science Foundation (NSF); the US Department of Energy Office of High-Energy Physics; the Laboratory Directed Research and Development (LDRD) program of Los Alamos National Laboratory; Consejo Nacional de Ciencia y Tecnolog\'{\'i}a (CONACyT), M{\'e}xico, grants 271051, 232656, 260378, 179588, 254964, 258865, 243290, 132197, A1-S-46288, A1-S-22784, c{\'a}tedras 873, 1563, 341, 323, Red HAWC, M{\'e}xico; DGAPA-UNAM grants AG100317, IN111315, IN111716-3, IN111419, IA102019, IN112218; VIEP-BUAP; PIFI 2012, 2013, PROFOCIE 2014, 2015; the University of Wisconsin Alumni Research Foundation; the Institute of Geophysics, Planetary Physics, and Signatures at Los Alamos National Laboratory; Polish Science Centre grant DEC-2018/31/B/ST9/01069, DEC-2017/27/B/ST9/02272; Coordinaci{\'o}n de la Investigaci{\'o}n Cient\'{\'i}fica de la Universidad Michoacana; Royal Society - Newton Advanced Fellowship 180385. Thanks to Scott Delay, Luciano D\'{\'i}az and Eduardo Murrieta for technical support. 
\software{threeML / hawc-hal (https://github.com/threeML, G. Vianello et. al.), AERIE-LiFF (https://github.com/rjlauer/aerie-liff, HAWC Collaboration)}

\bibliography{references}

\clearpage
\begin{appendix}

  \section*{Results}

  \begin{figure}
      \centering
      \includegraphics[width=.32\textwidth]{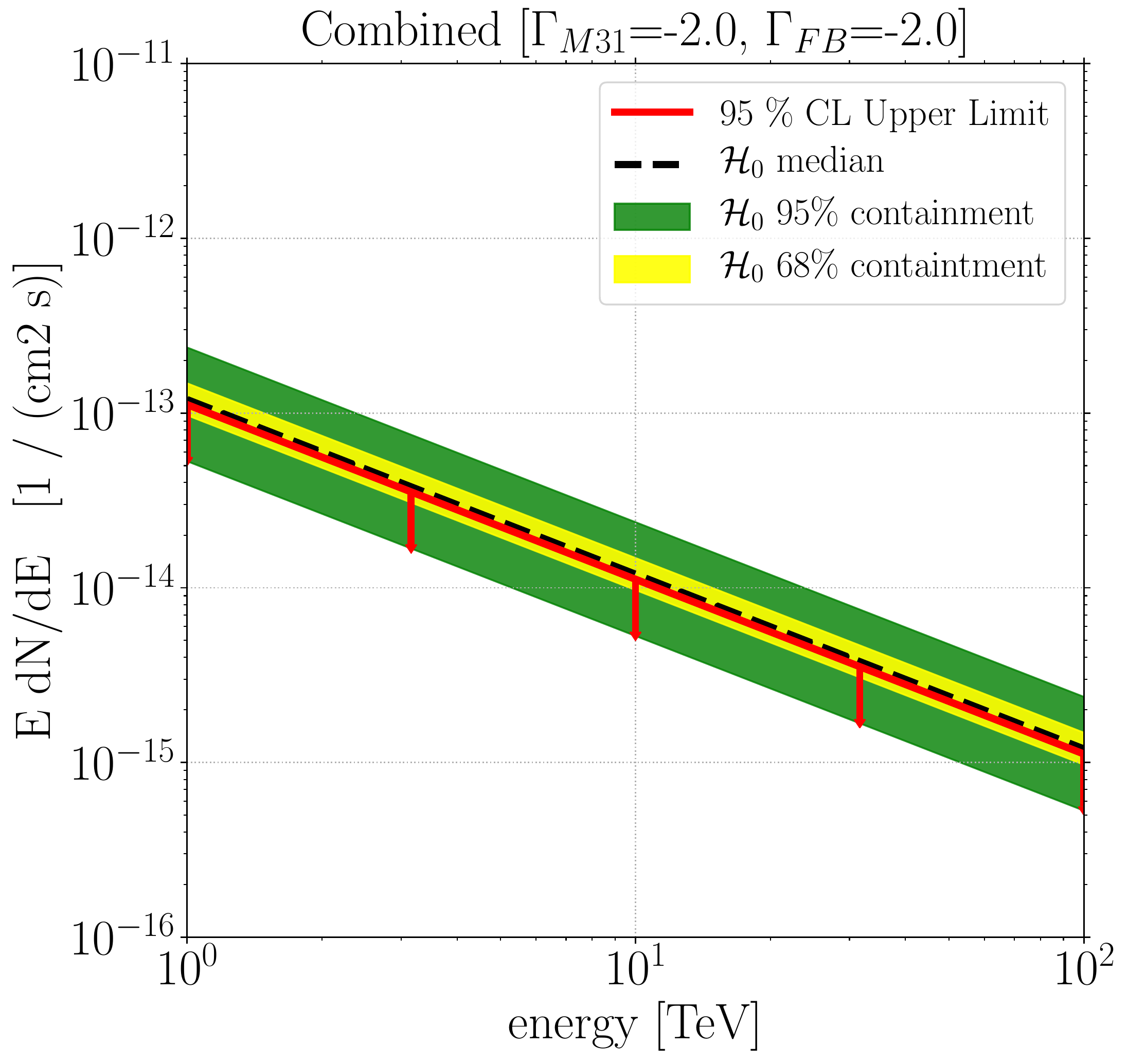}
      \hfill
      \includegraphics[width=.32\textwidth]{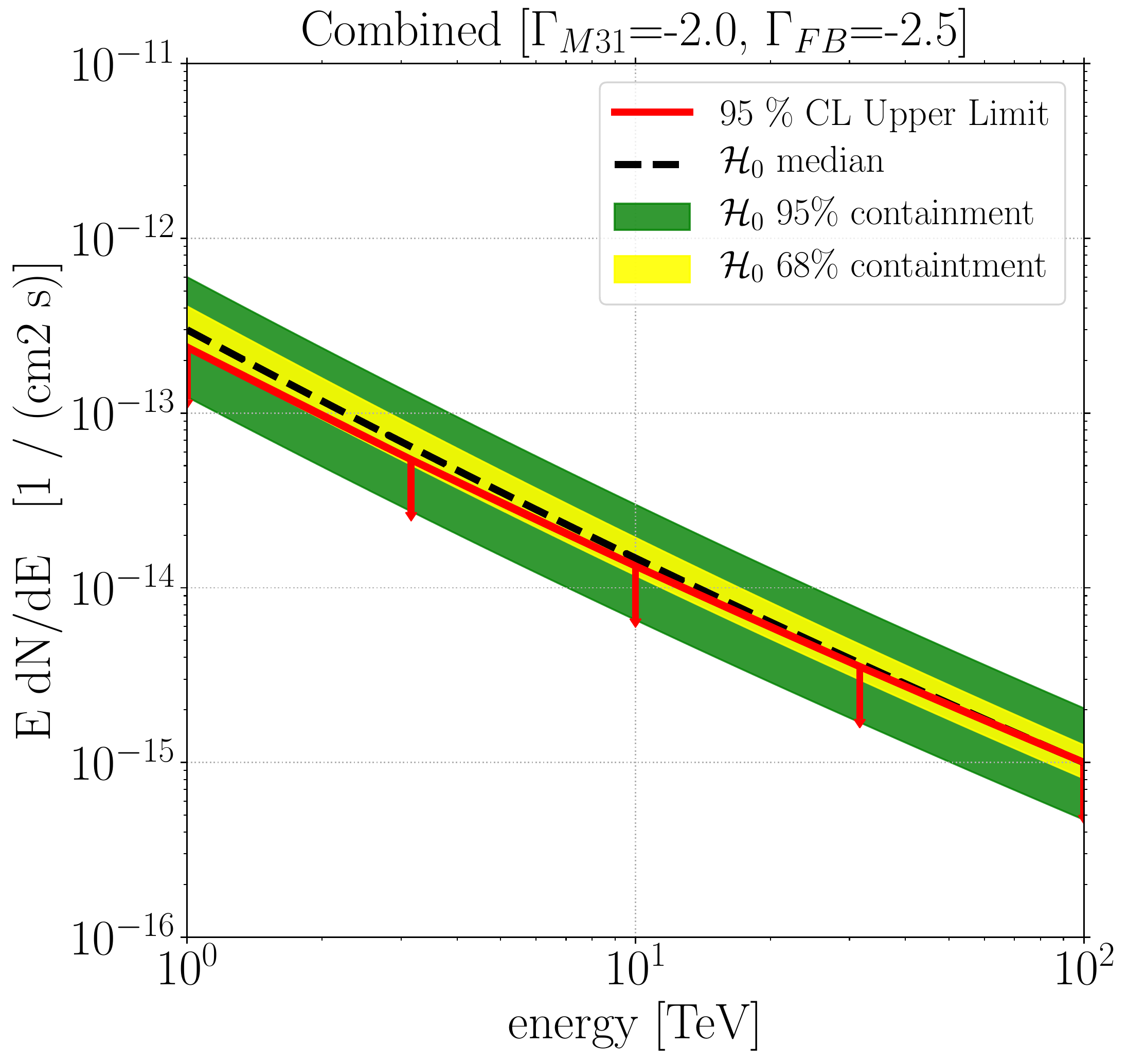}
      \hfill
      \includegraphics[width=.32\textwidth]{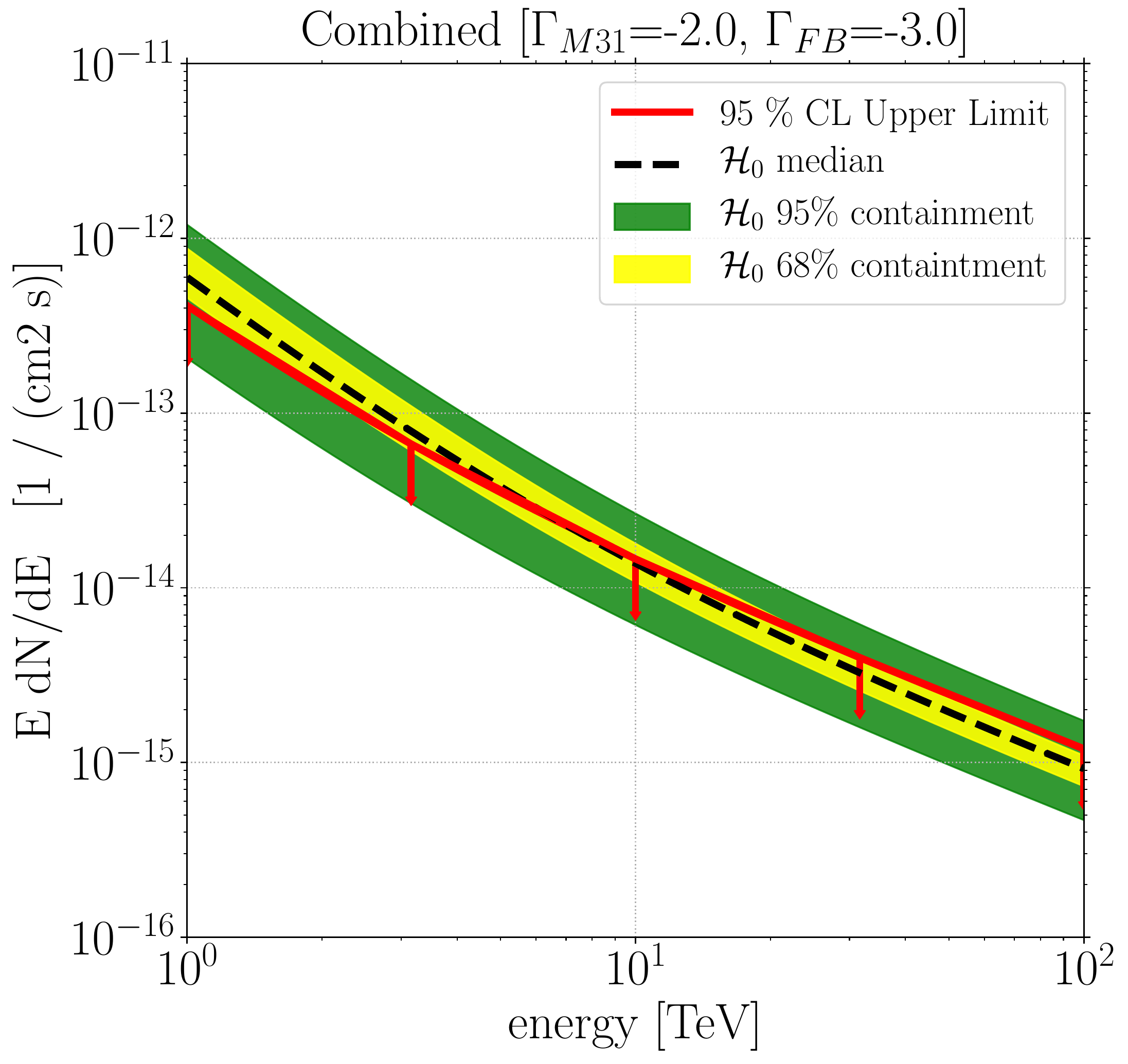}\\
      \includegraphics[width=.32\textwidth]{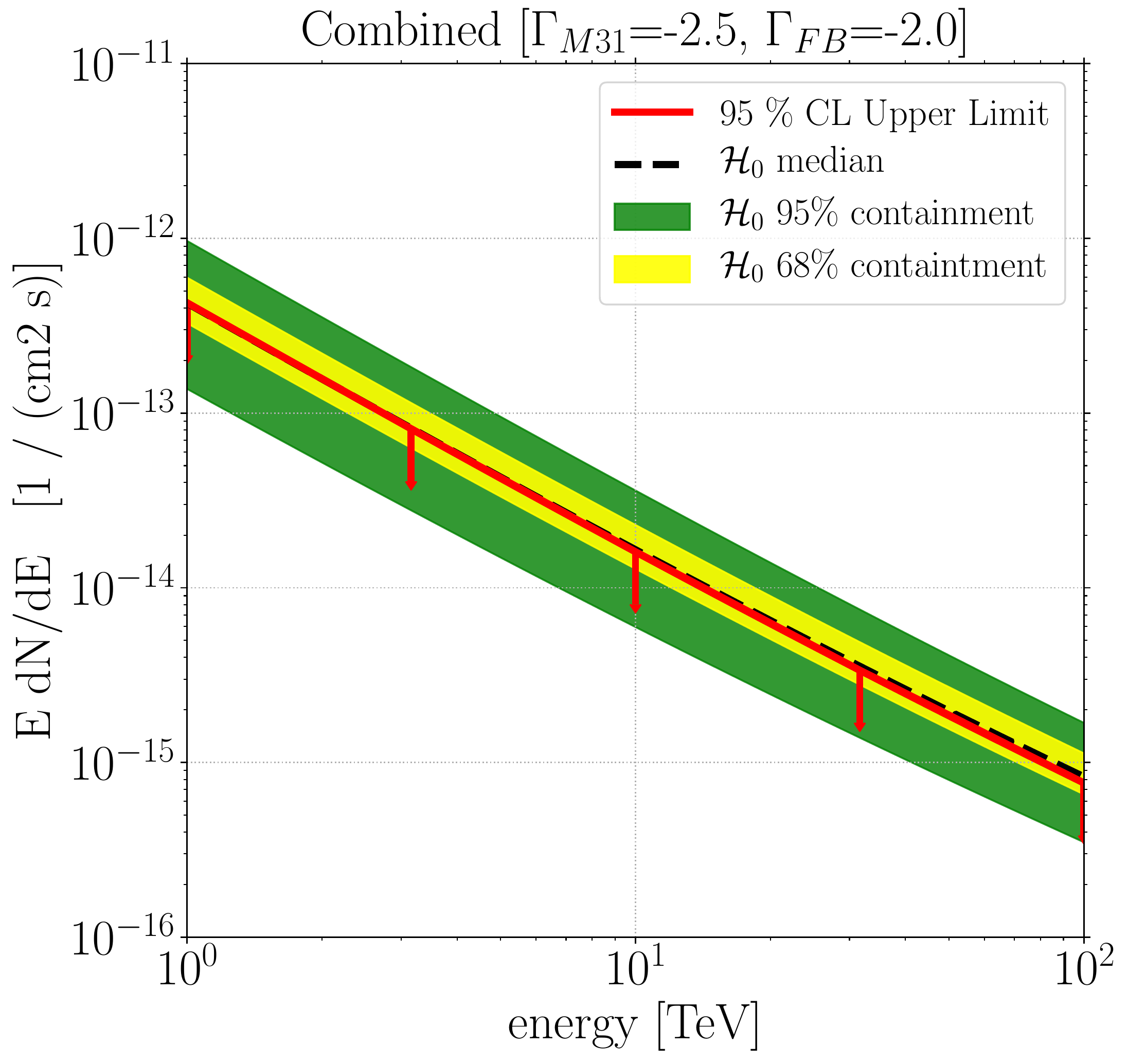}
      \hfill
      \includegraphics[width=.32\textwidth]{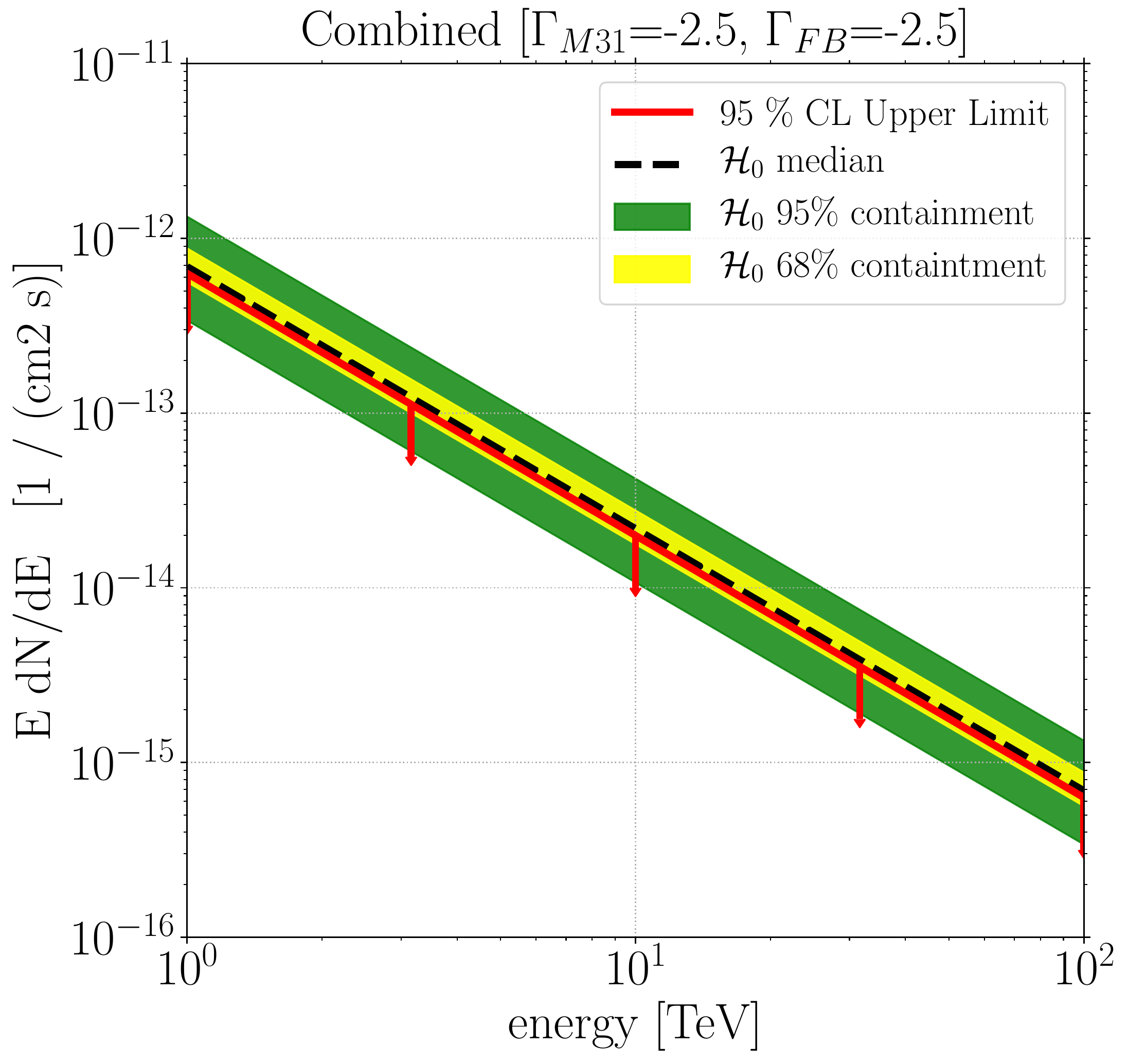}
      \hfill
      \includegraphics[width=.32\textwidth]{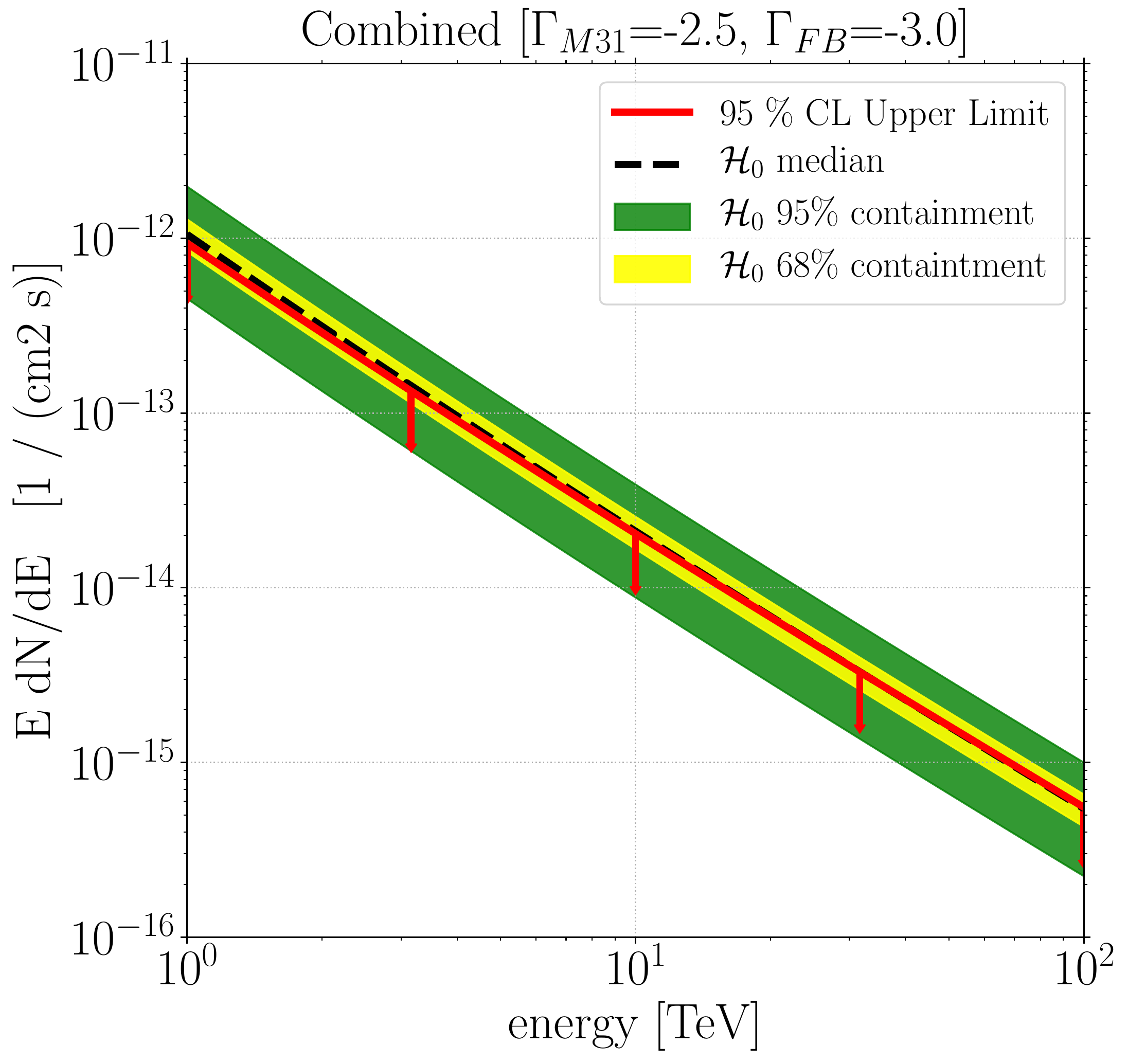}\\
      \includegraphics[width=.32\textwidth]{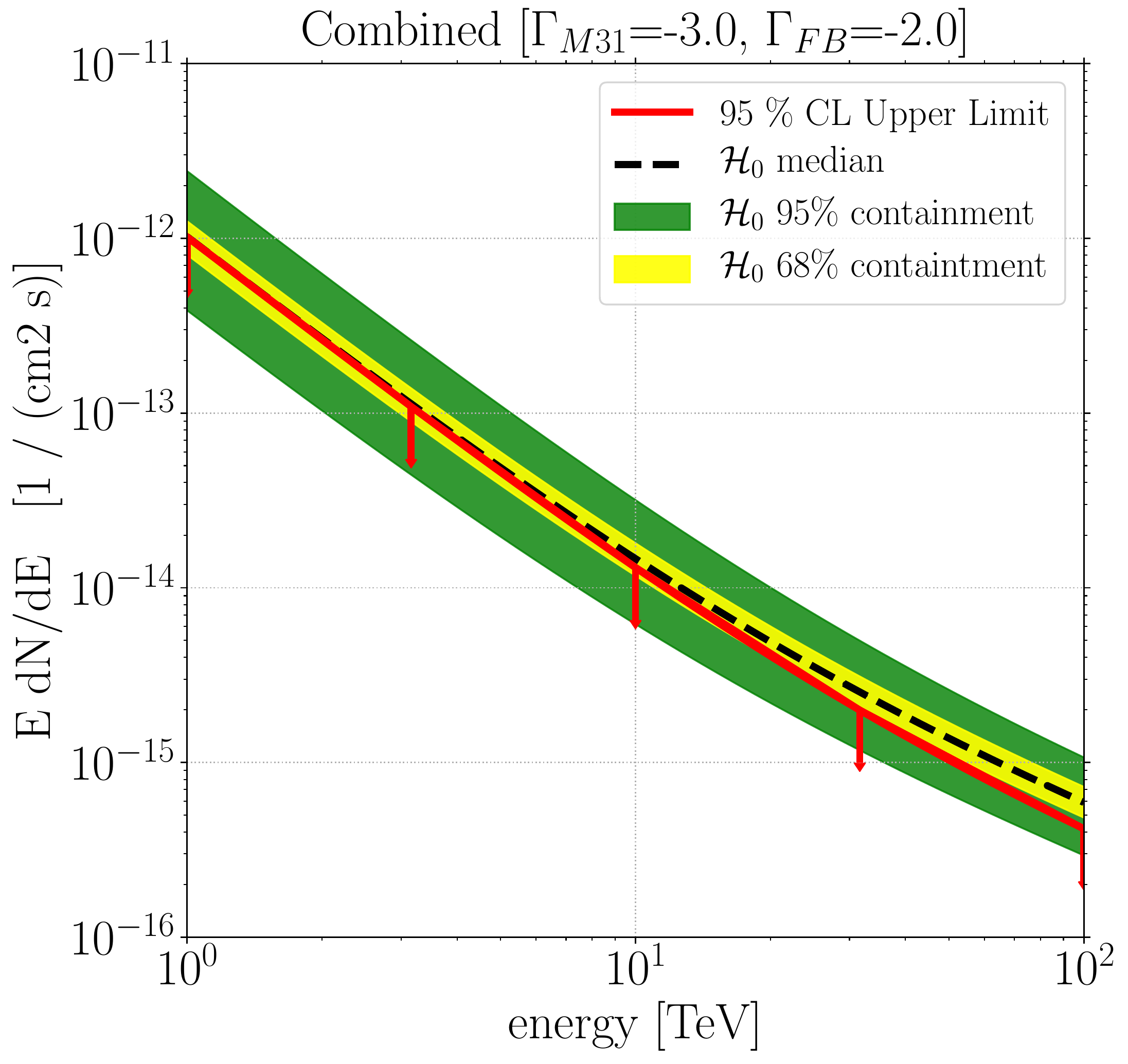}
      \hfill
      \includegraphics[width=.32\textwidth]{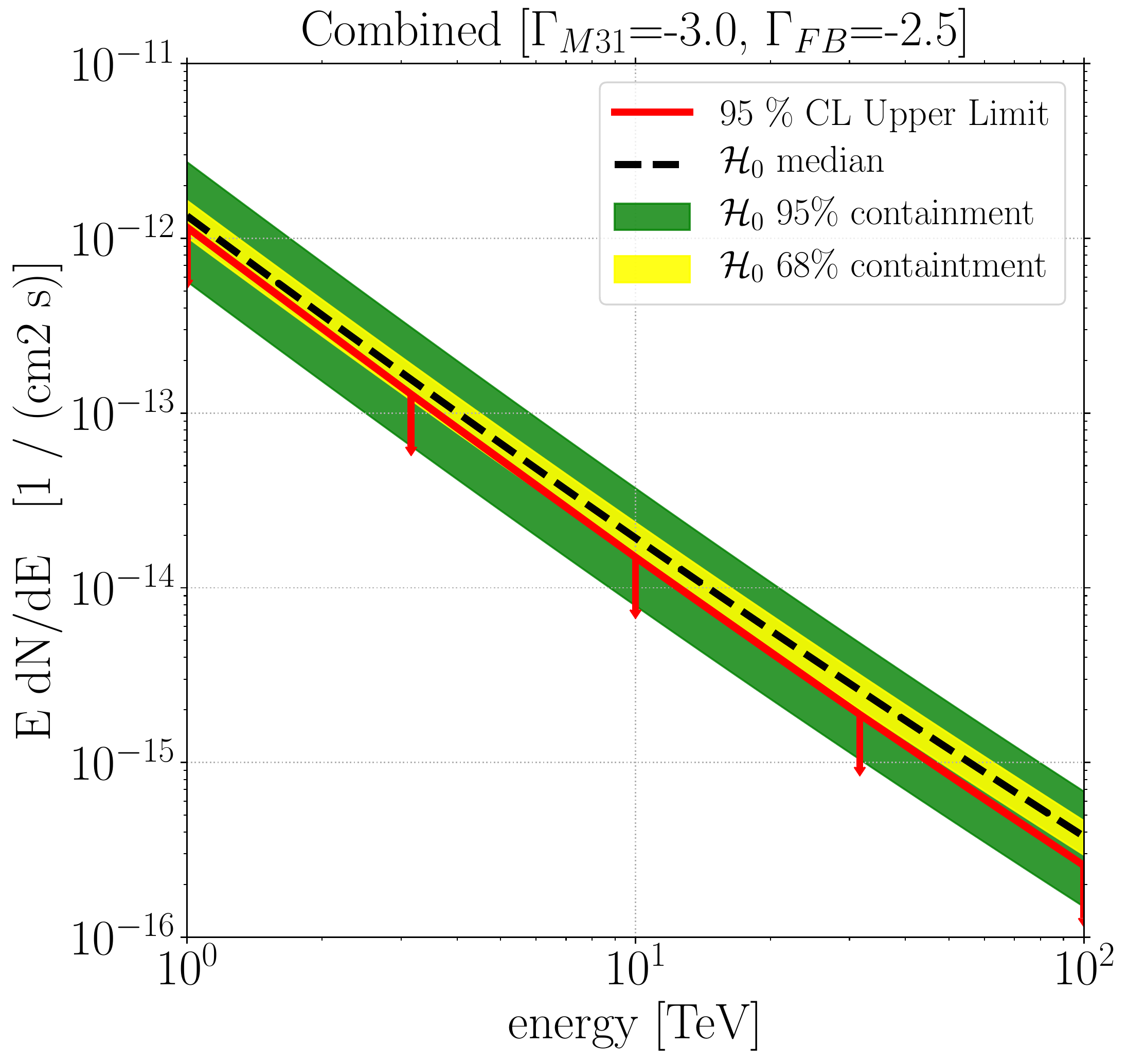}
      \hfill
      \includegraphics[width=.32\textwidth]{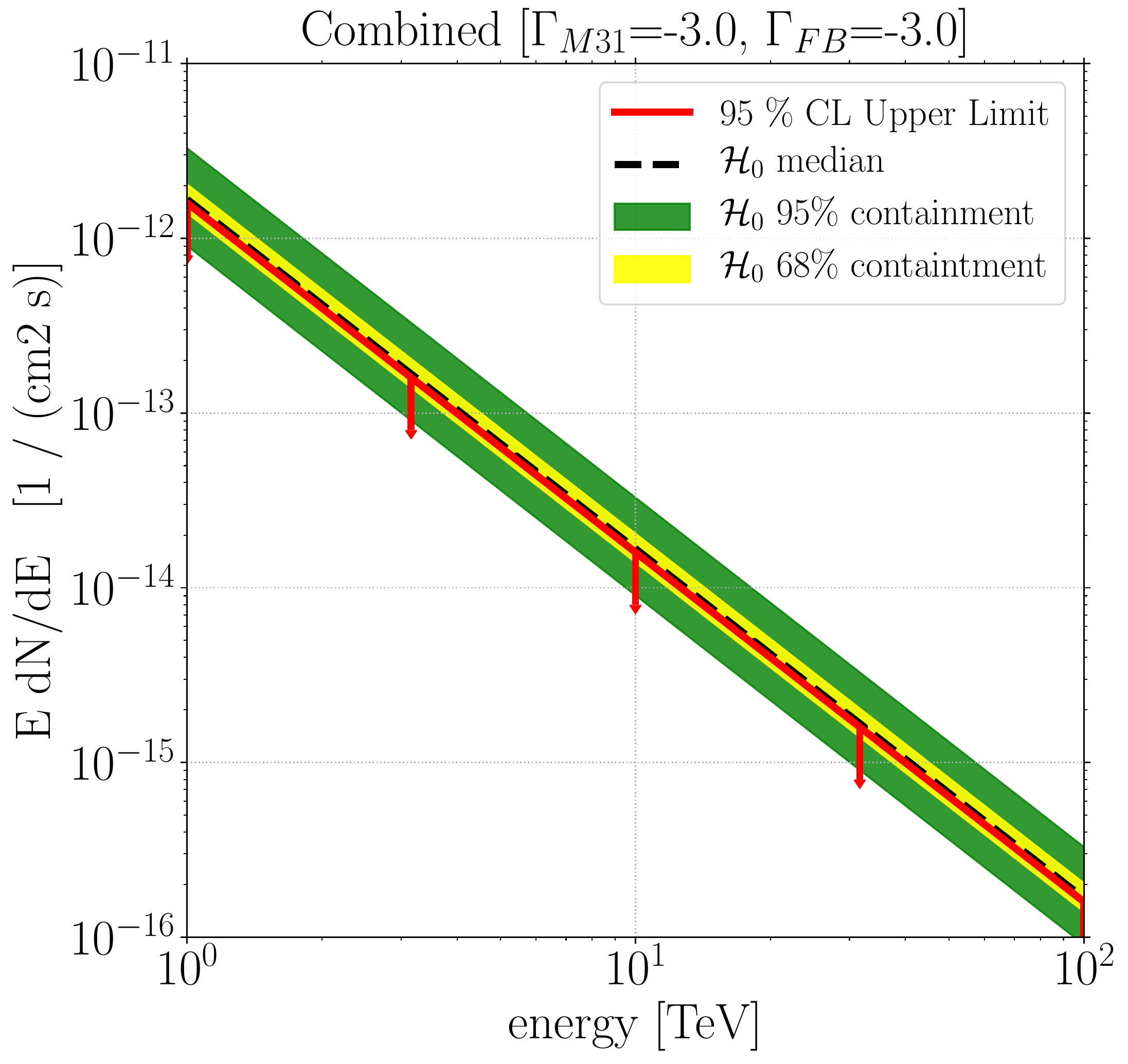}\\
      \caption{\label{fig:combined} 95\% credible interval upper intervals on the gamma-ray emission from the M31 disk and the M31 FBs, with nine combinations of spectral indices for the two regions.}
  \end{figure}

  \begin{figure}
      \centering
      \includegraphics[width=.5\textwidth]{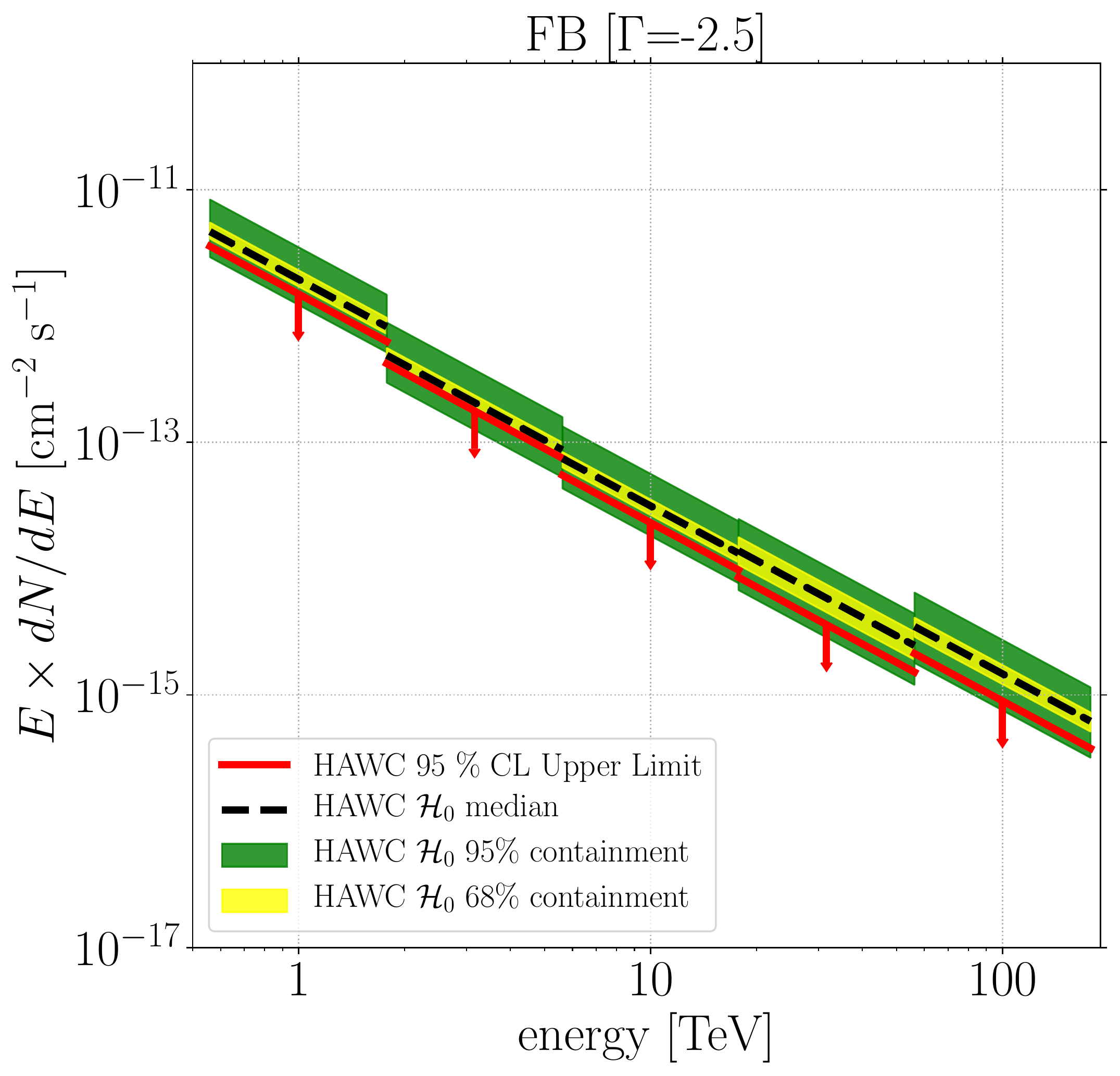}
      \caption{Measured quasi-differential upper limits and expected limits for the M31 FBs for a spectral index of 2.5.}
      \label{fig:quasi_bubbles_appendix}
  \end{figure}

  \begin{figure}
      \centering
      \includegraphics[width=.32\textwidth]{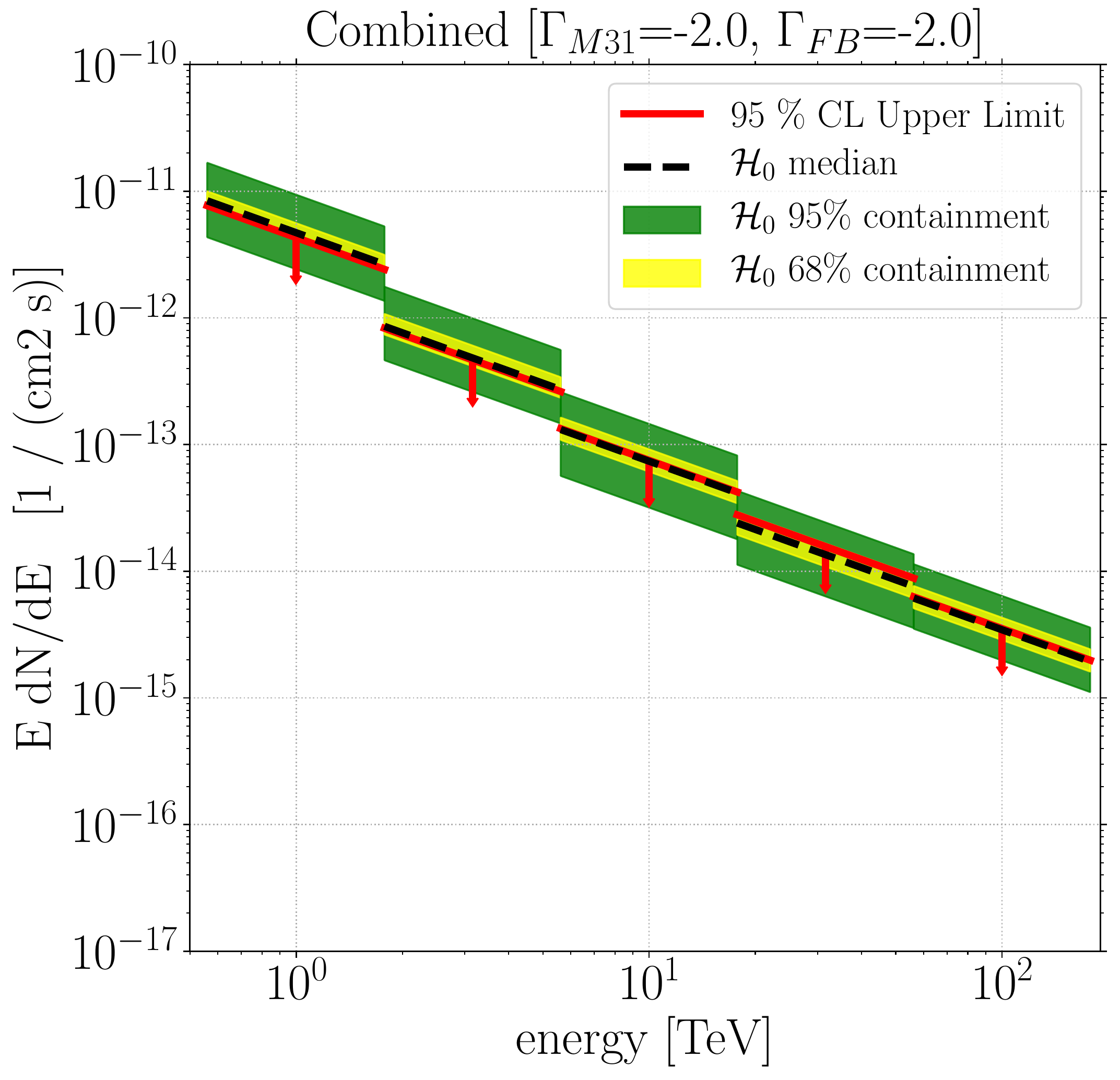}
      \includegraphics[width=.32\textwidth]{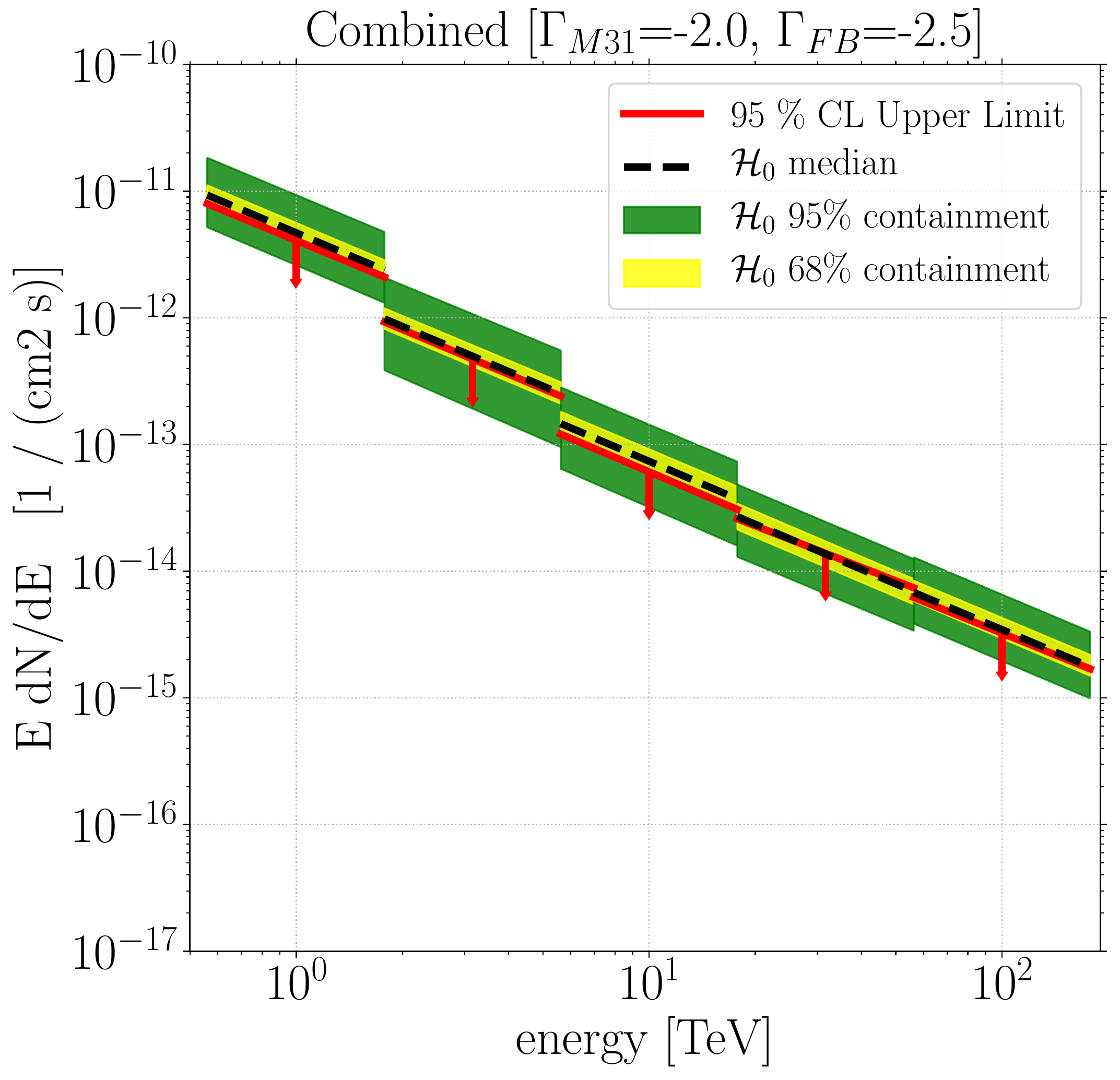}
      \includegraphics[width=.32\textwidth]{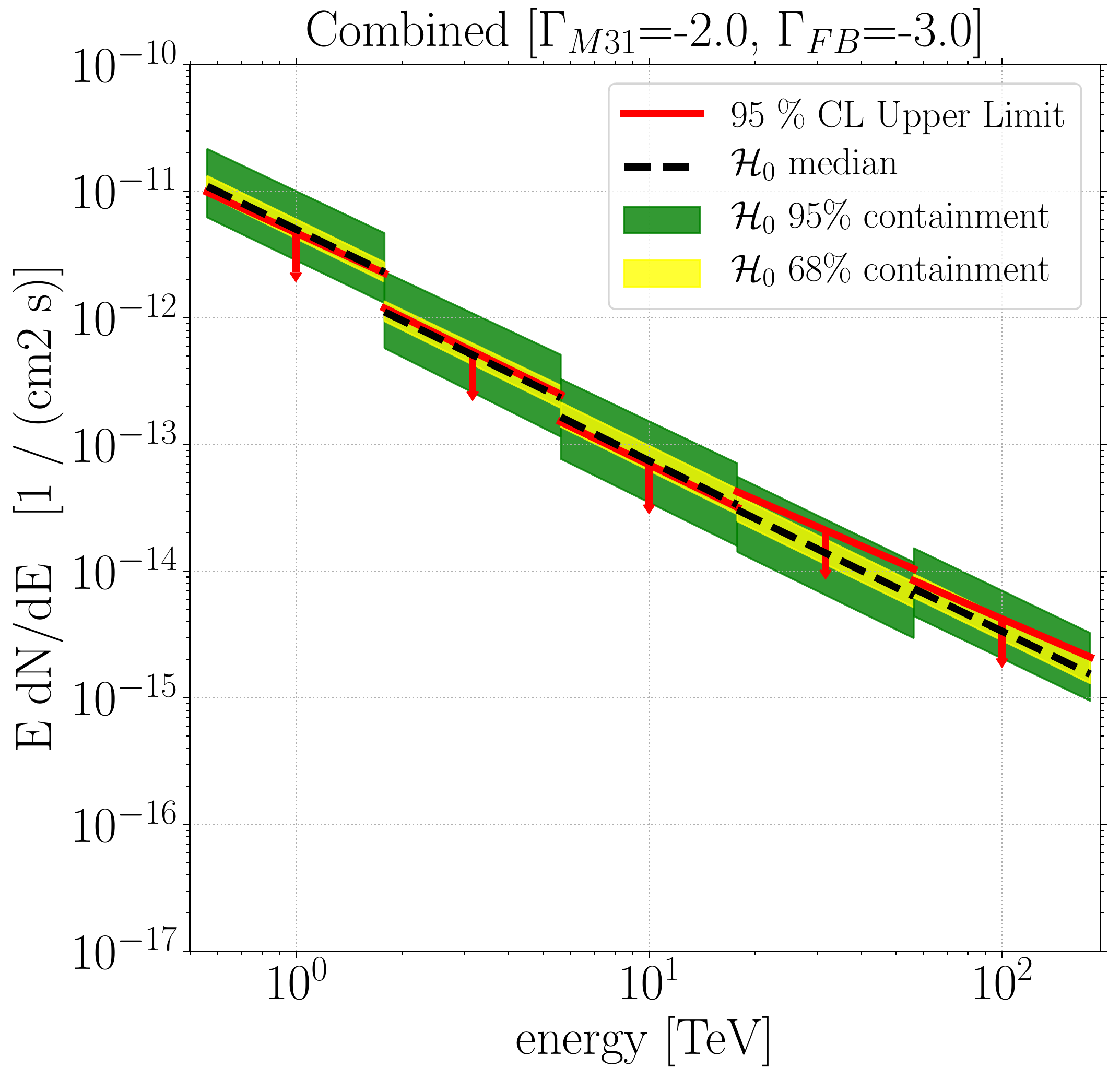}
      \includegraphics[width=.32\textwidth]{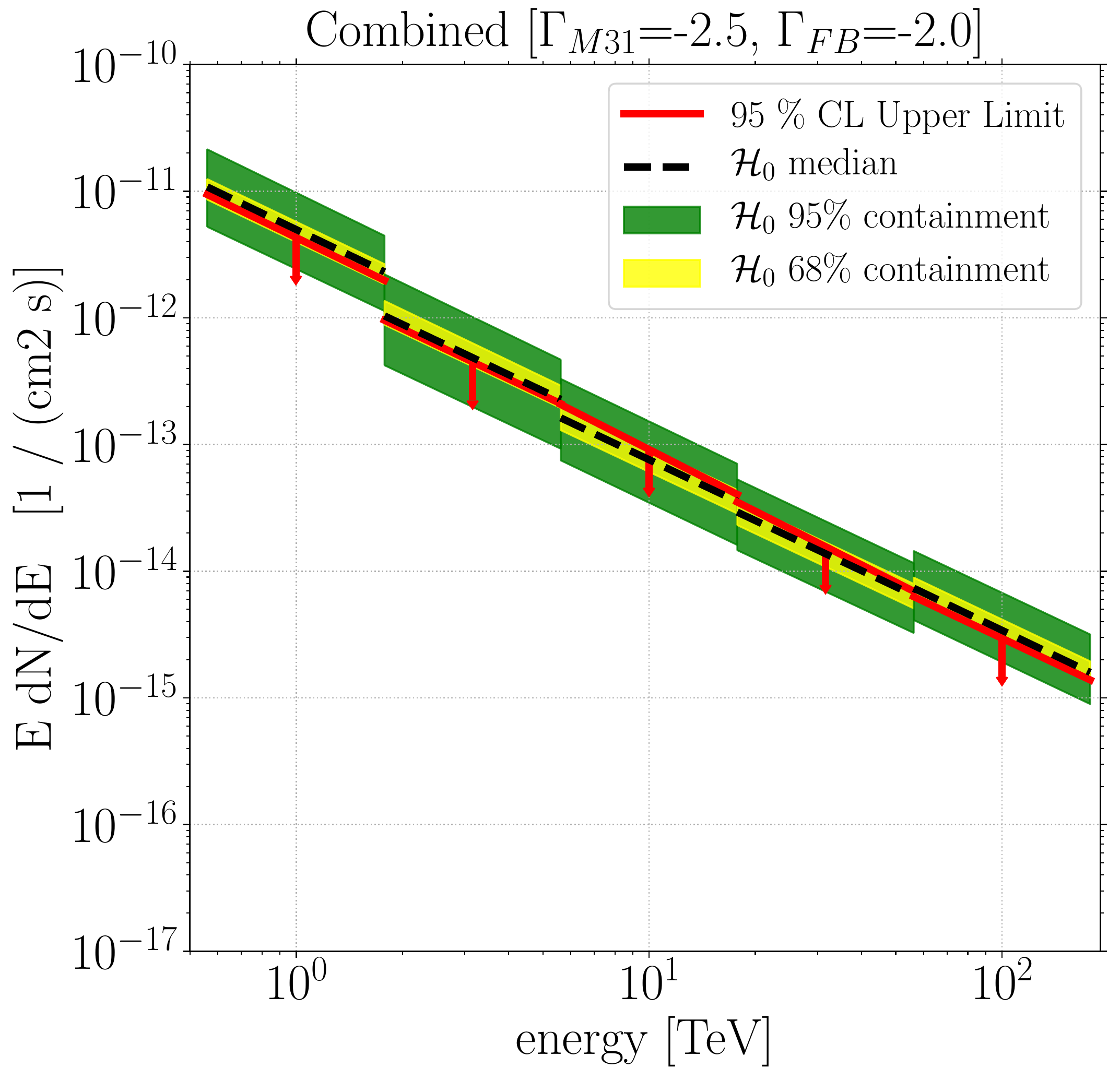}
      \includegraphics[width=.32\textwidth]{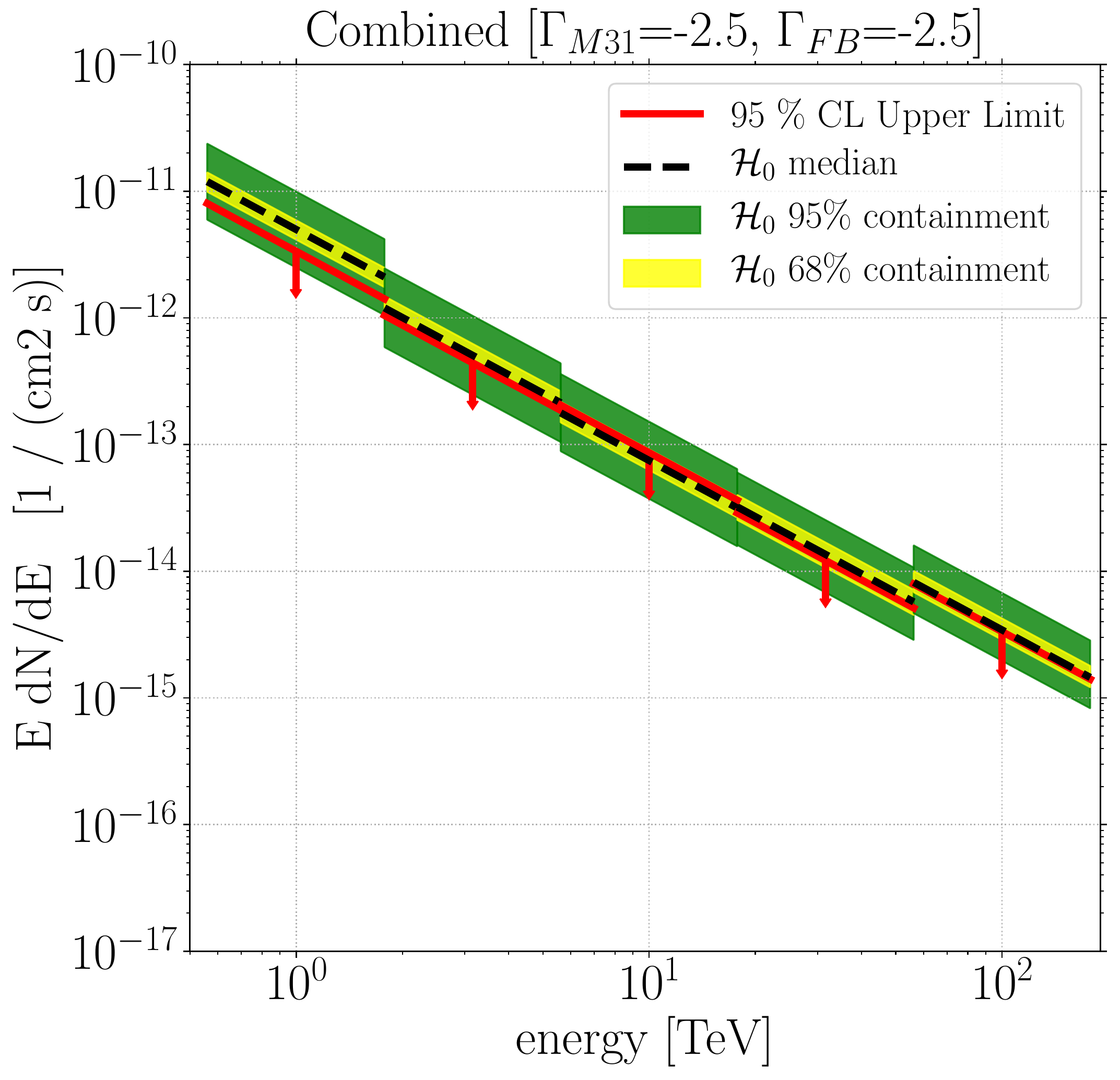}
      \includegraphics[width=.32\textwidth]{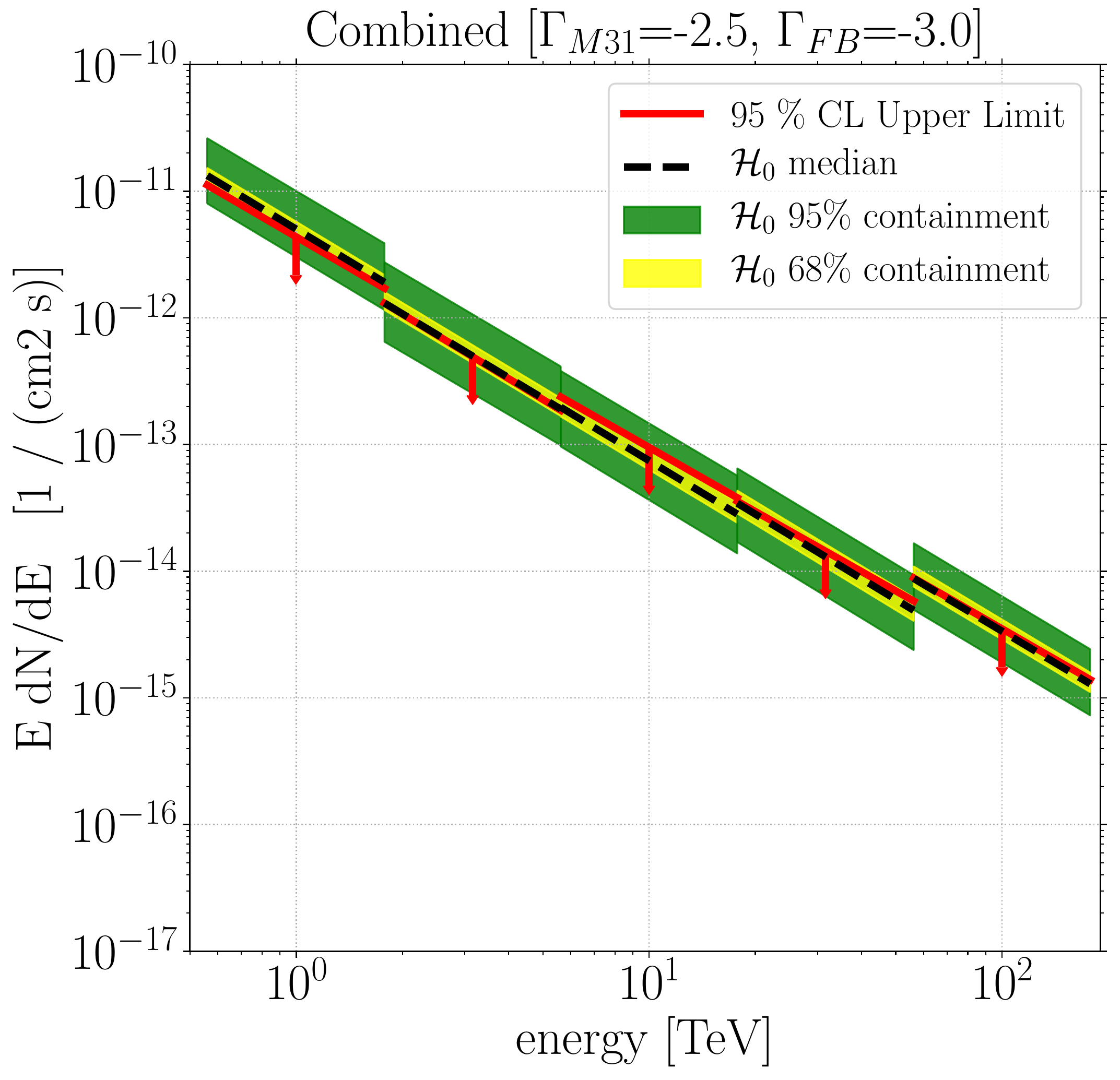}
      \includegraphics[width=.32\textwidth]{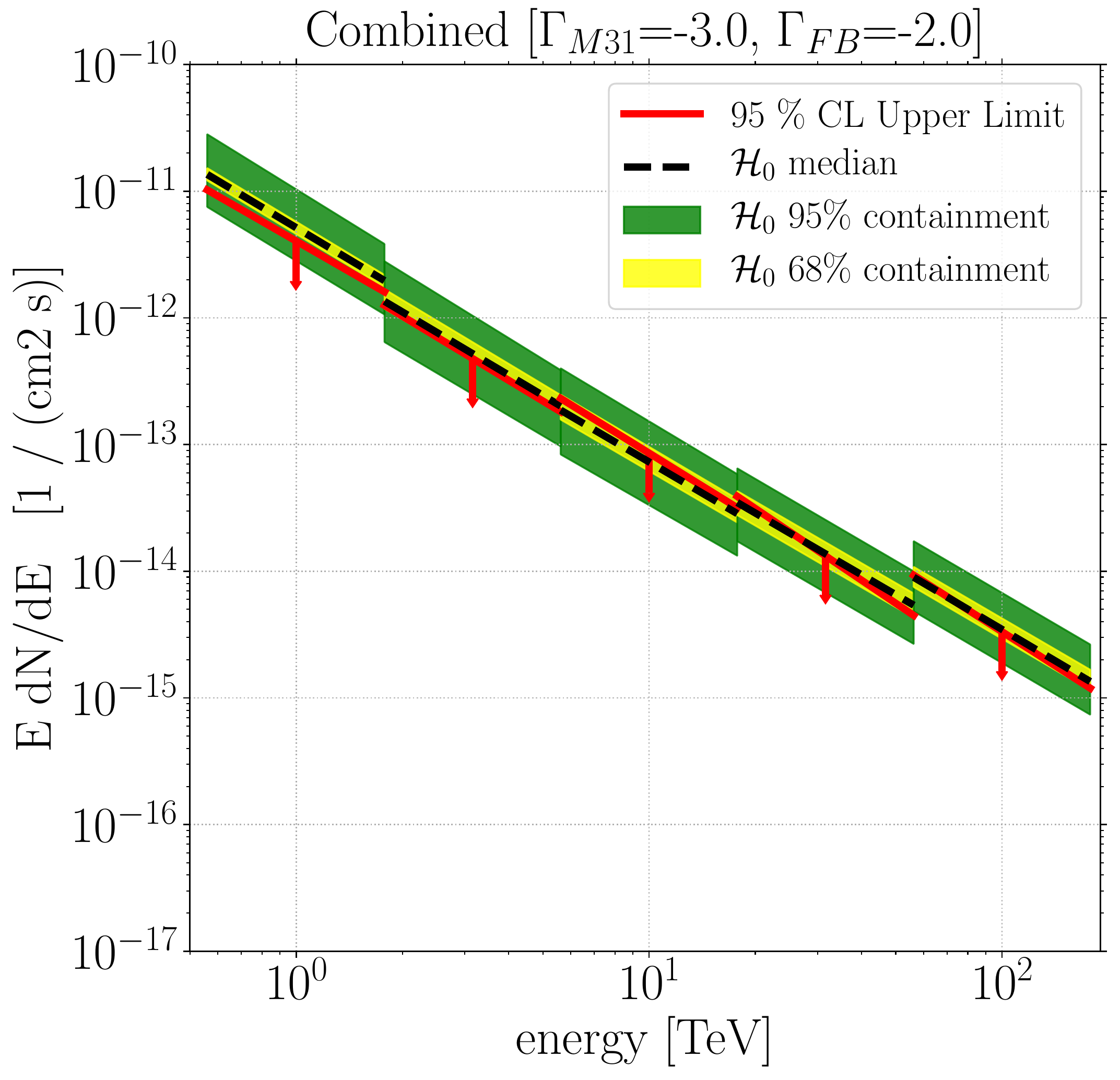}
      \includegraphics[width=.32\textwidth]{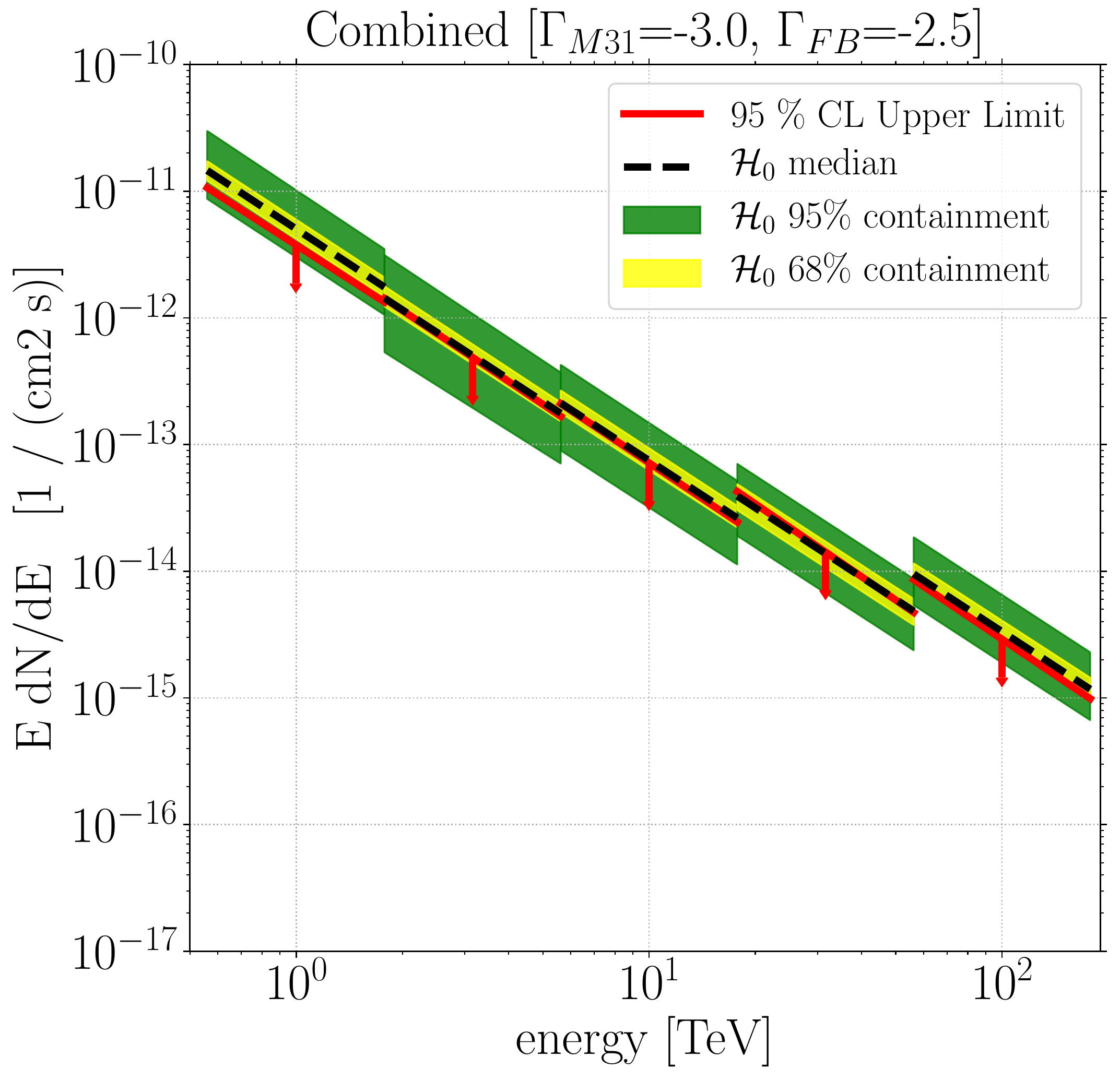}
      \includegraphics[width=.32\textwidth]{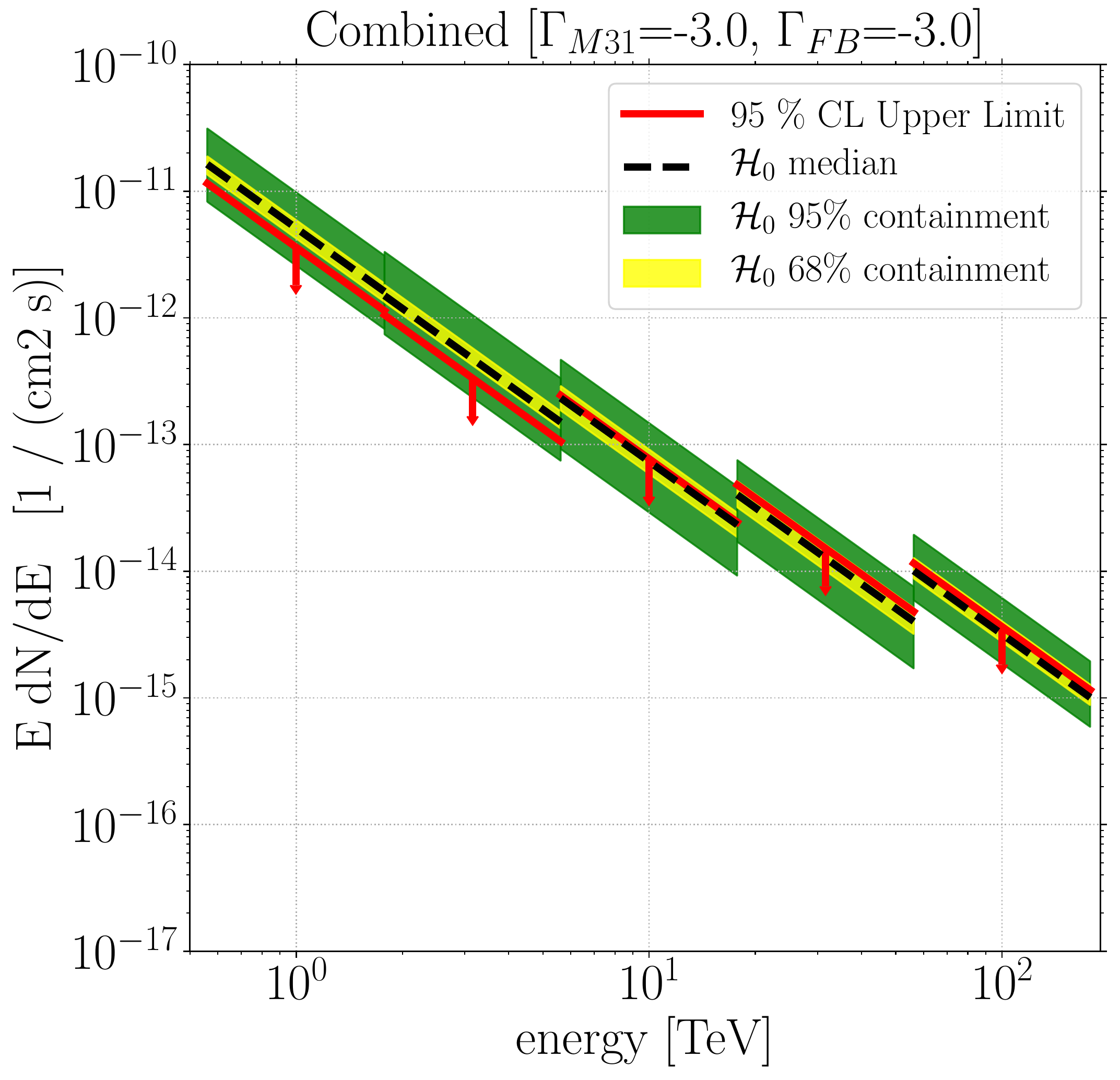}
      \caption{Measured quasi-differential upper limits and expected limits for the M31 disk and the FBs for different spectral index assumptions.}
      \label{fig:quasi_combined2}
  \end{figure}

  \begin{figure}[h!]
    \centering
    \includegraphics[width=.32\textwidth]{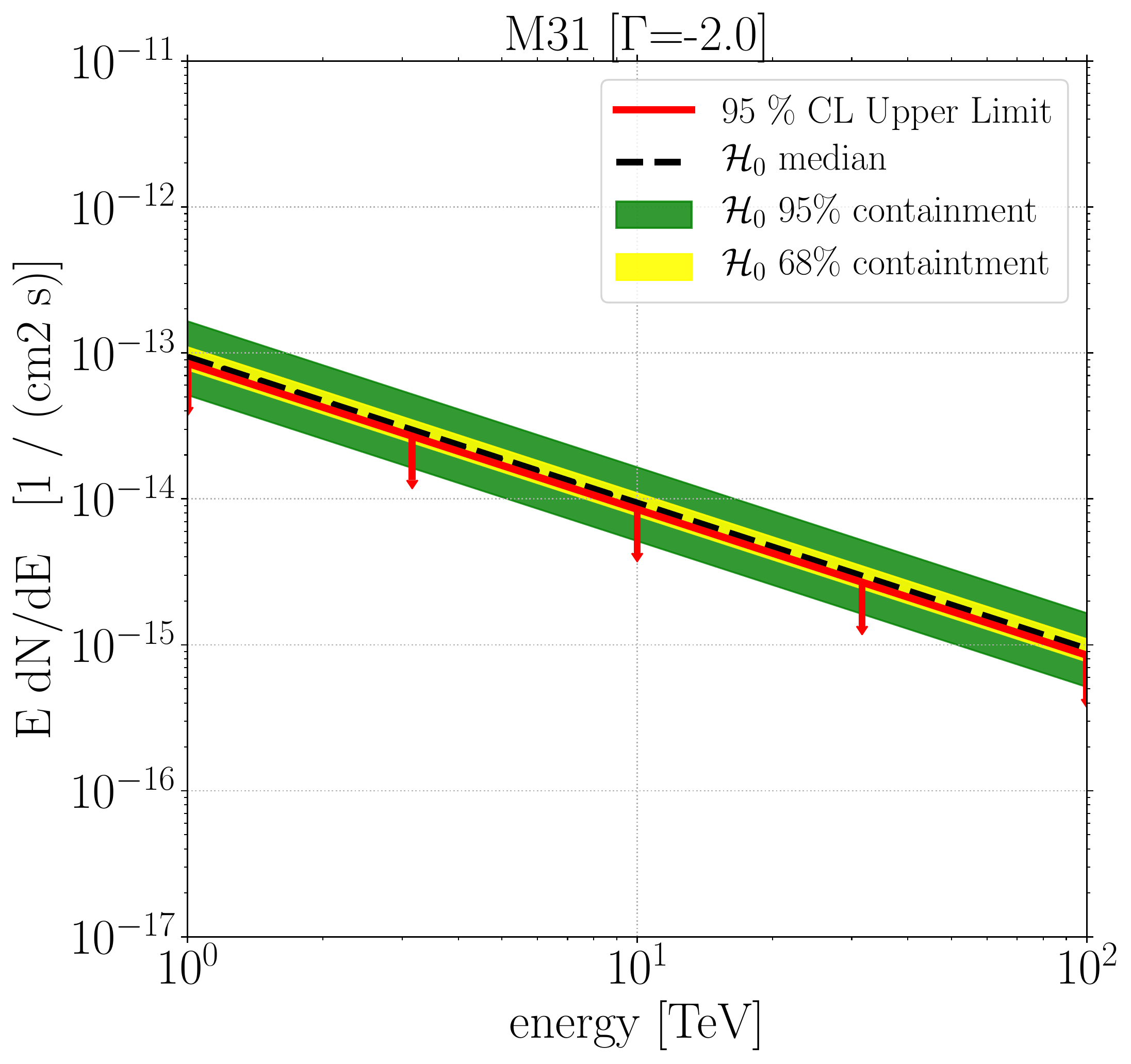}
    \hfill
    \includegraphics[width=.32\textwidth]{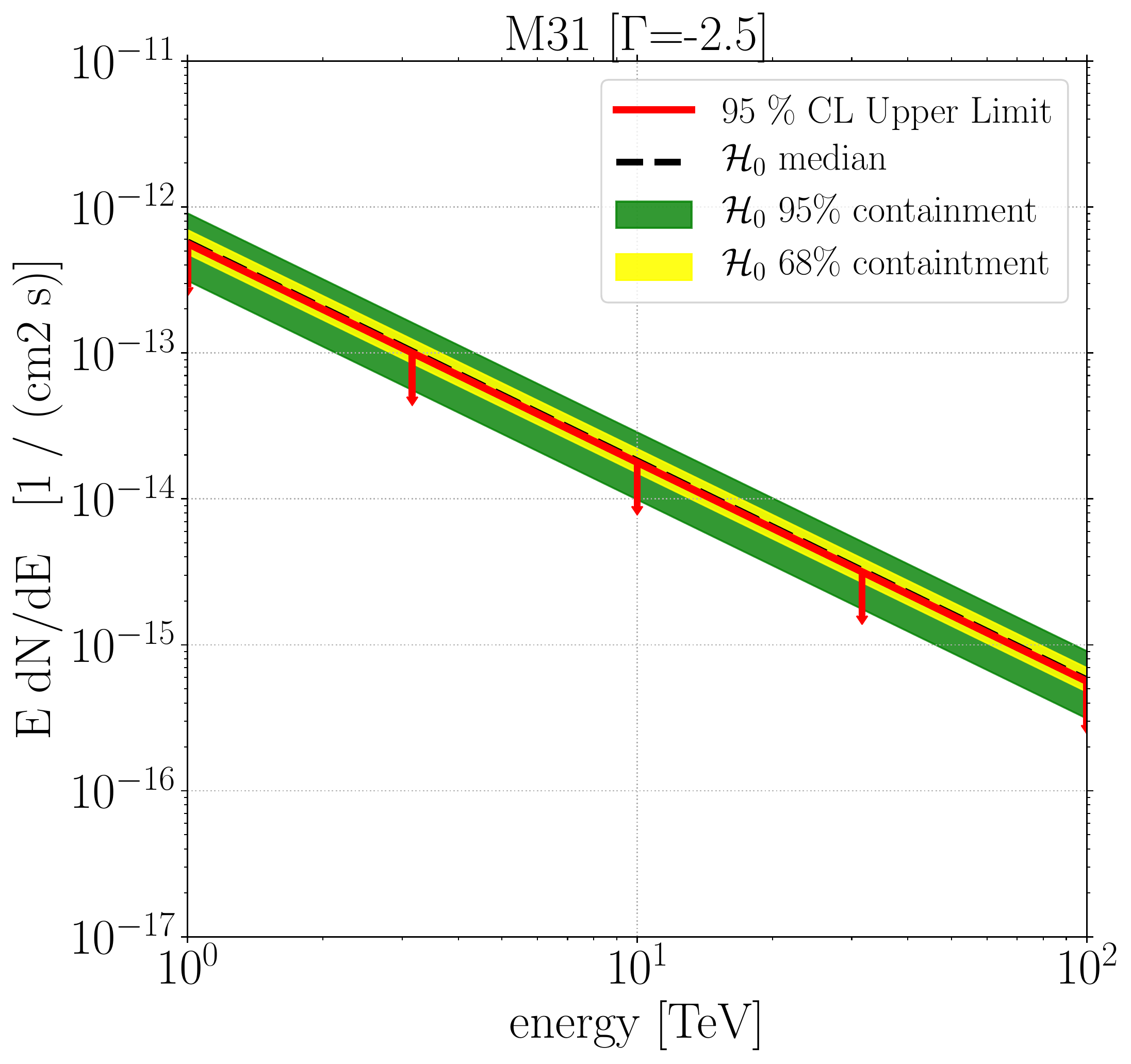}
    \hfill
    \includegraphics[width=.32\textwidth]{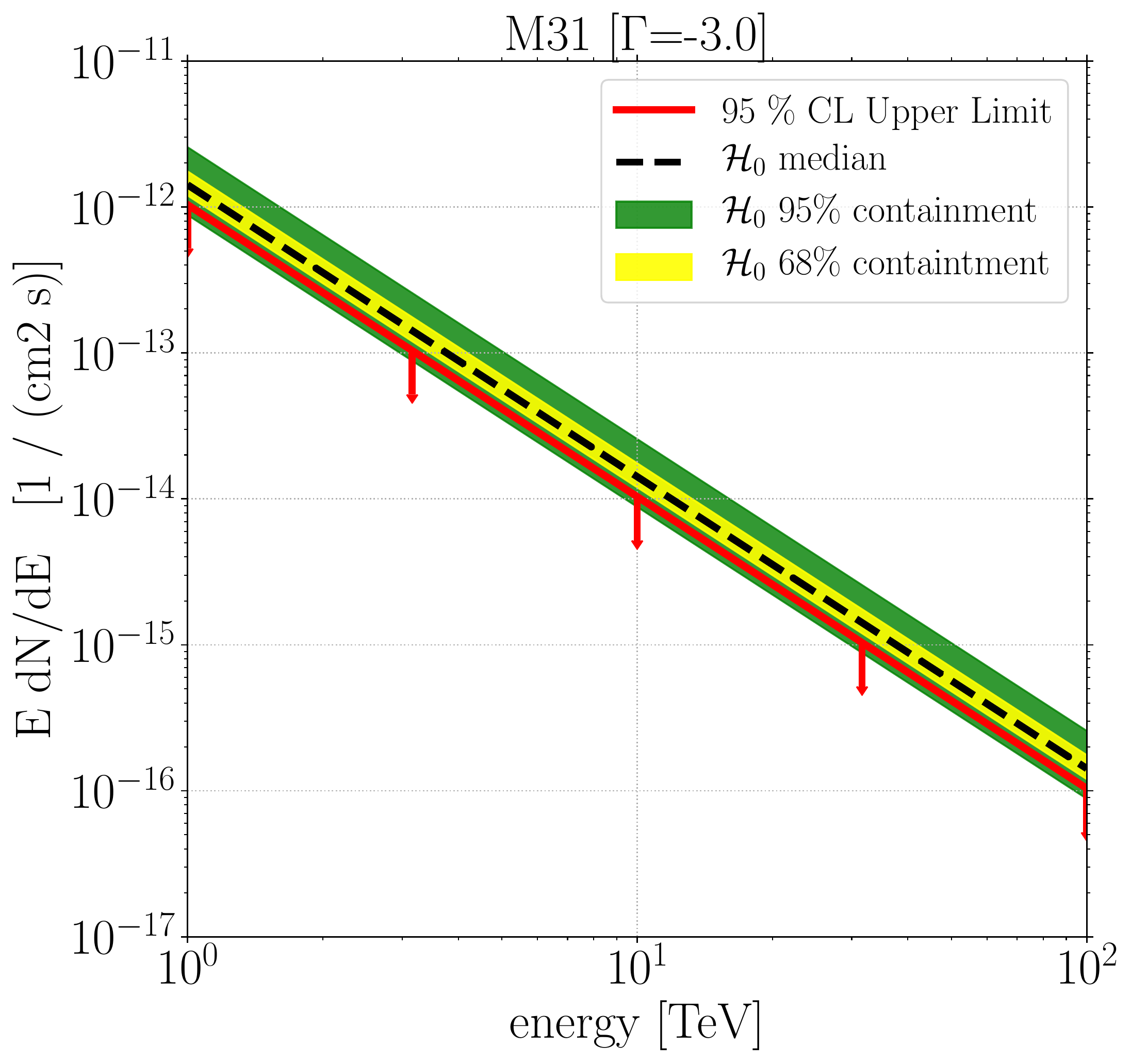}
    \caption{\label{fig:m31_results_appendix} 95\% credible interval upper intervals (and background-only expected upper limit bands) of the M31 disk assuming power-law emission with spectral index $-2$, $-2.5$, and $-3$. Dotted lines are the extrapolations of our upper limit for energies lower than HAWC's sensitivity.}
  \end{figure}

  \begin{figure}[h!]
    \centering
    \includegraphics[width=.32\textwidth]{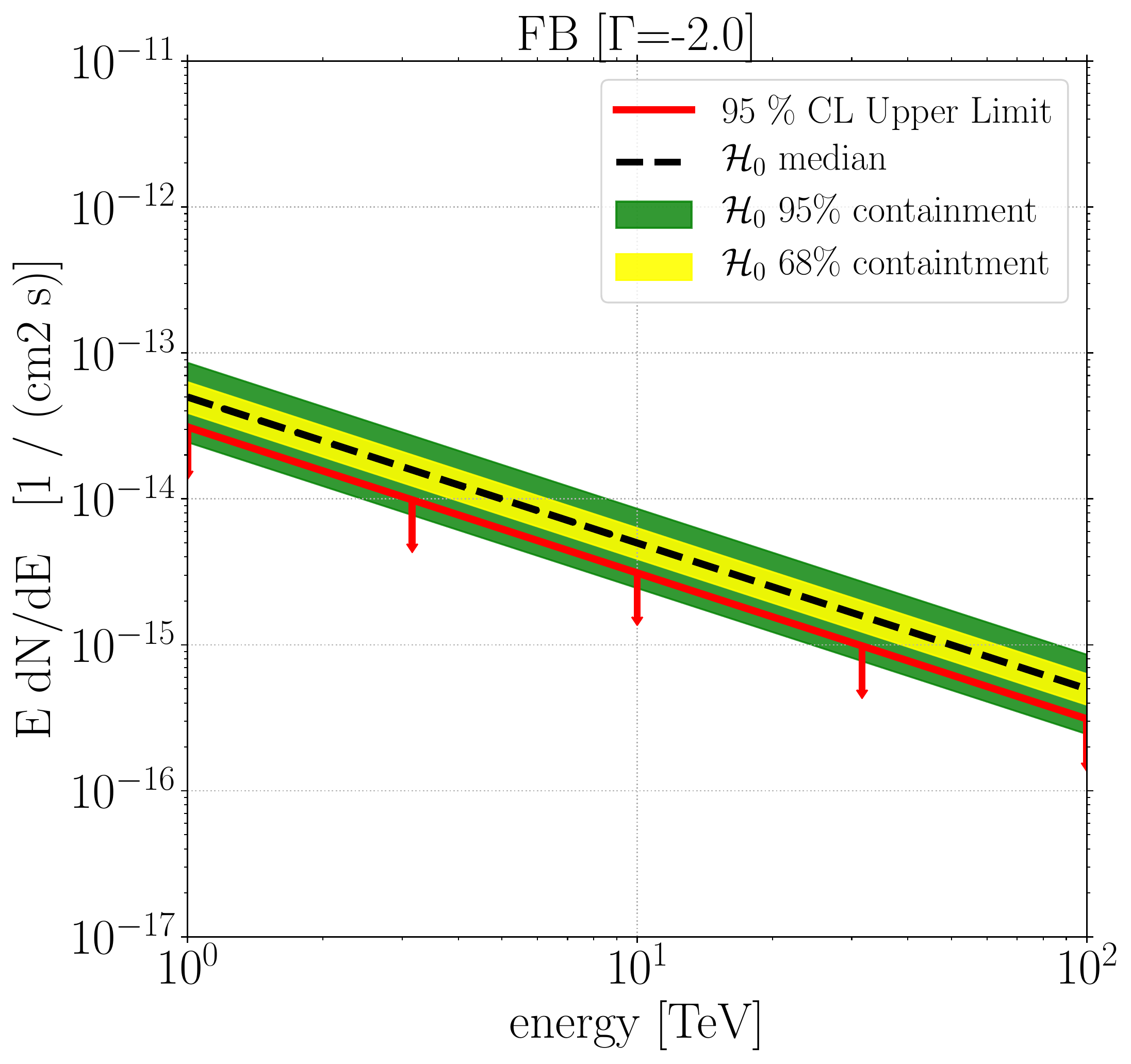}
    \hfill
    \includegraphics[width=.32\textwidth]{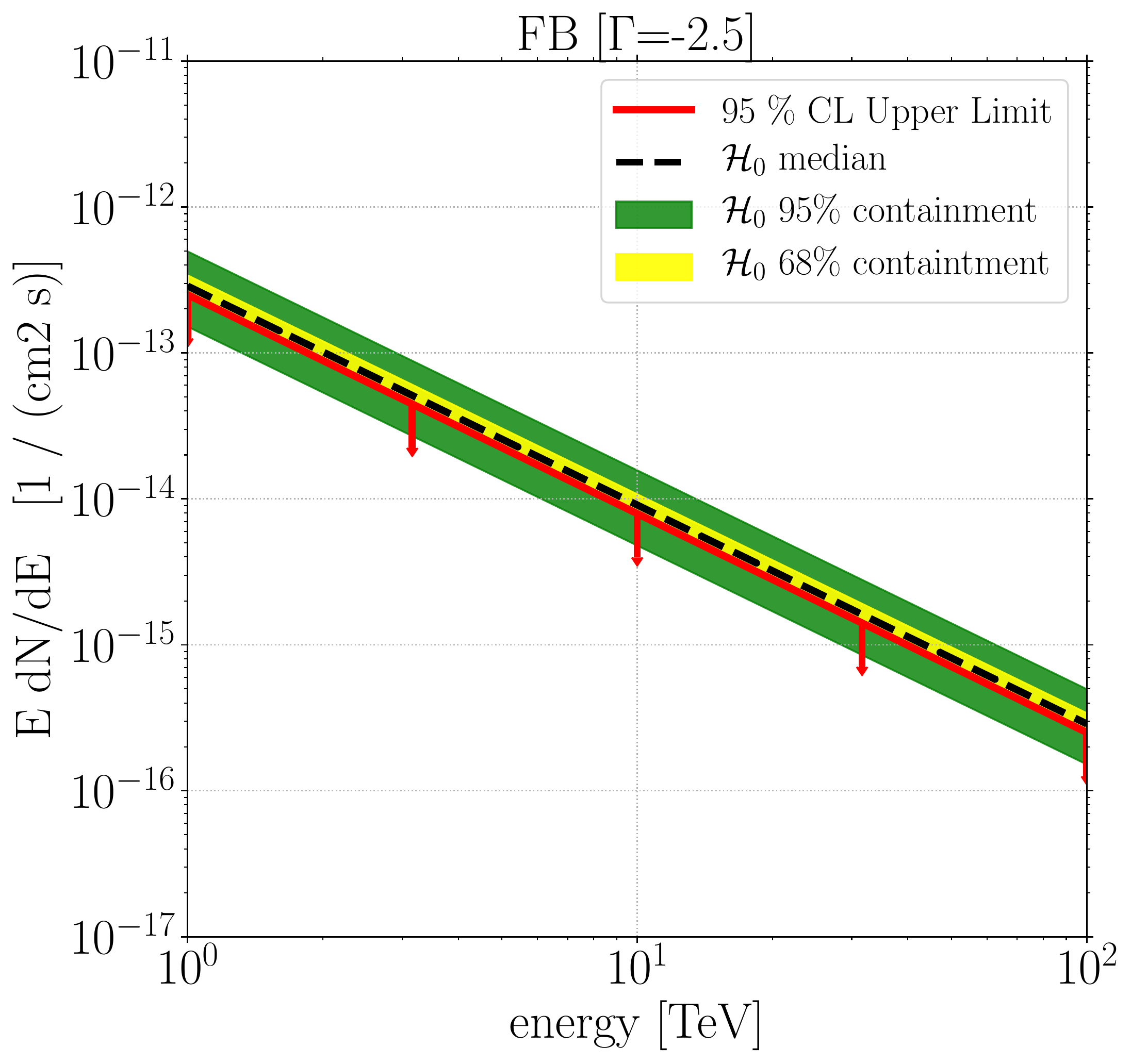}
    \hfill
    \includegraphics[width=.32\textwidth]{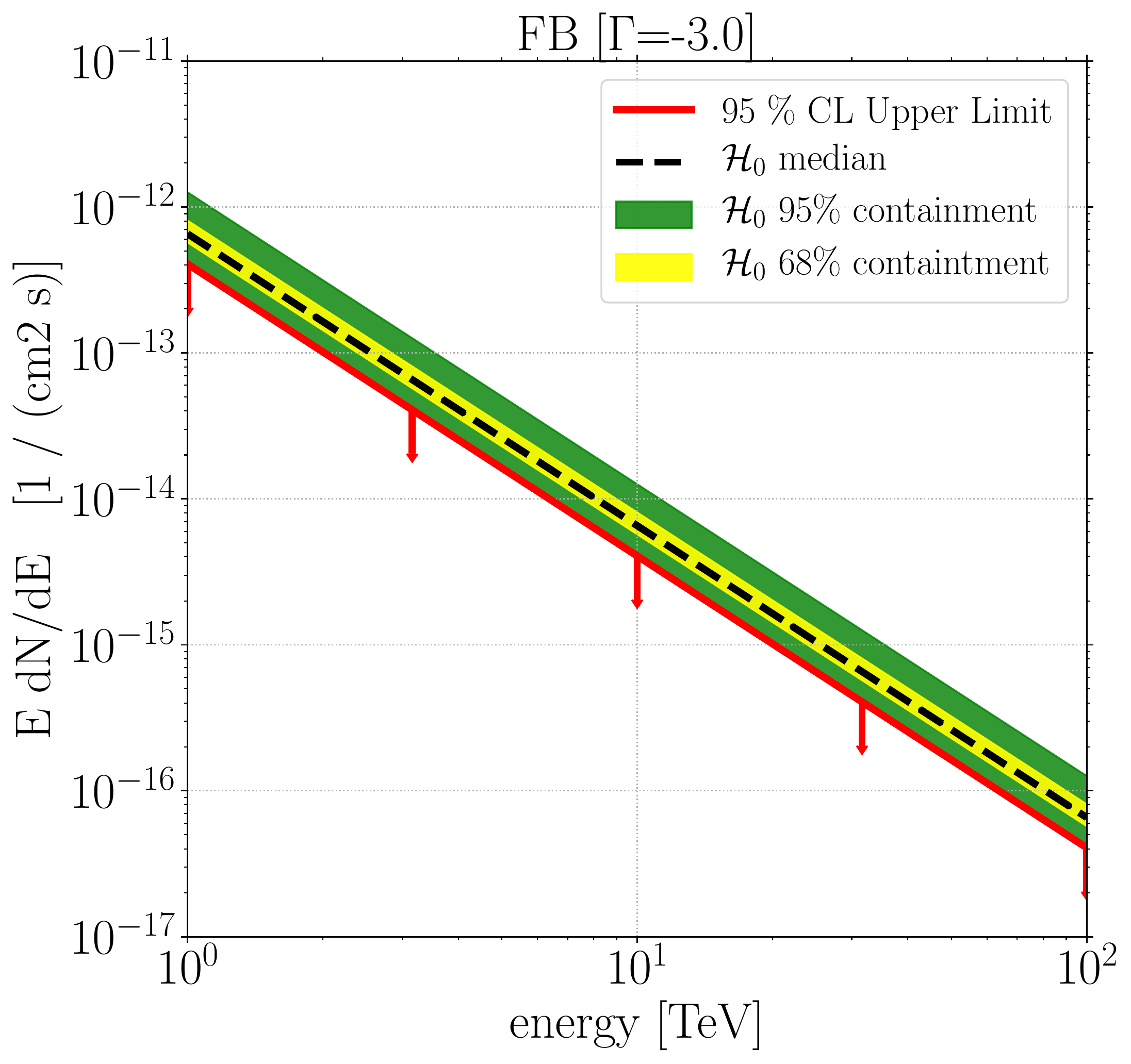}
    \caption{\label{fig:fb_results_appendix} 95\% credible interval upper limits and expected upper limits bands for the M31 ``bubble'' templates assuming a power law with spectral index $-2$, $-2.5$, and $-3$.}
\end{figure}

  \begin{figure}[h!]
    \centering
    \includegraphics[width=.32\textwidth]{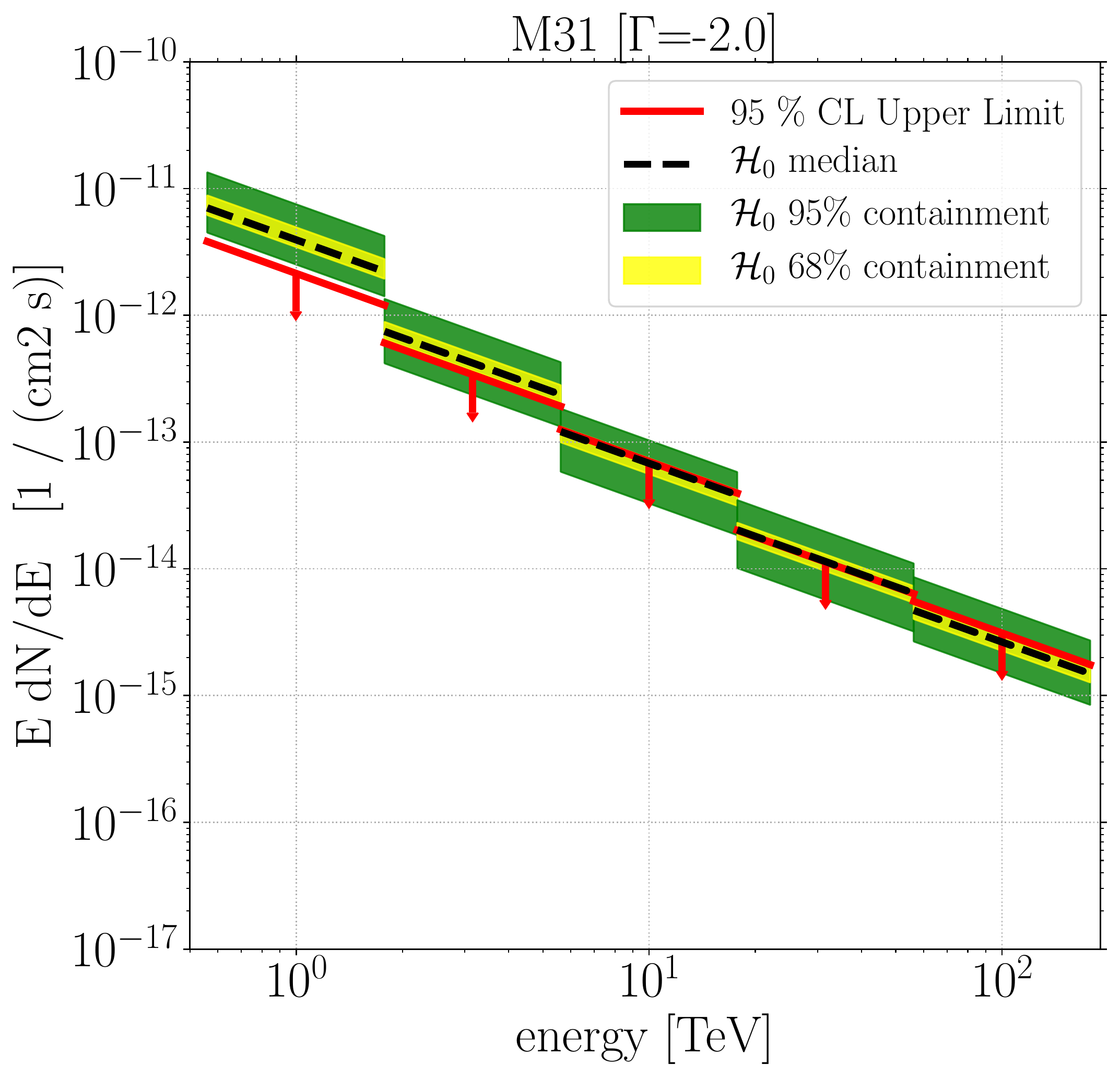}
    \includegraphics[width=.32\textwidth]{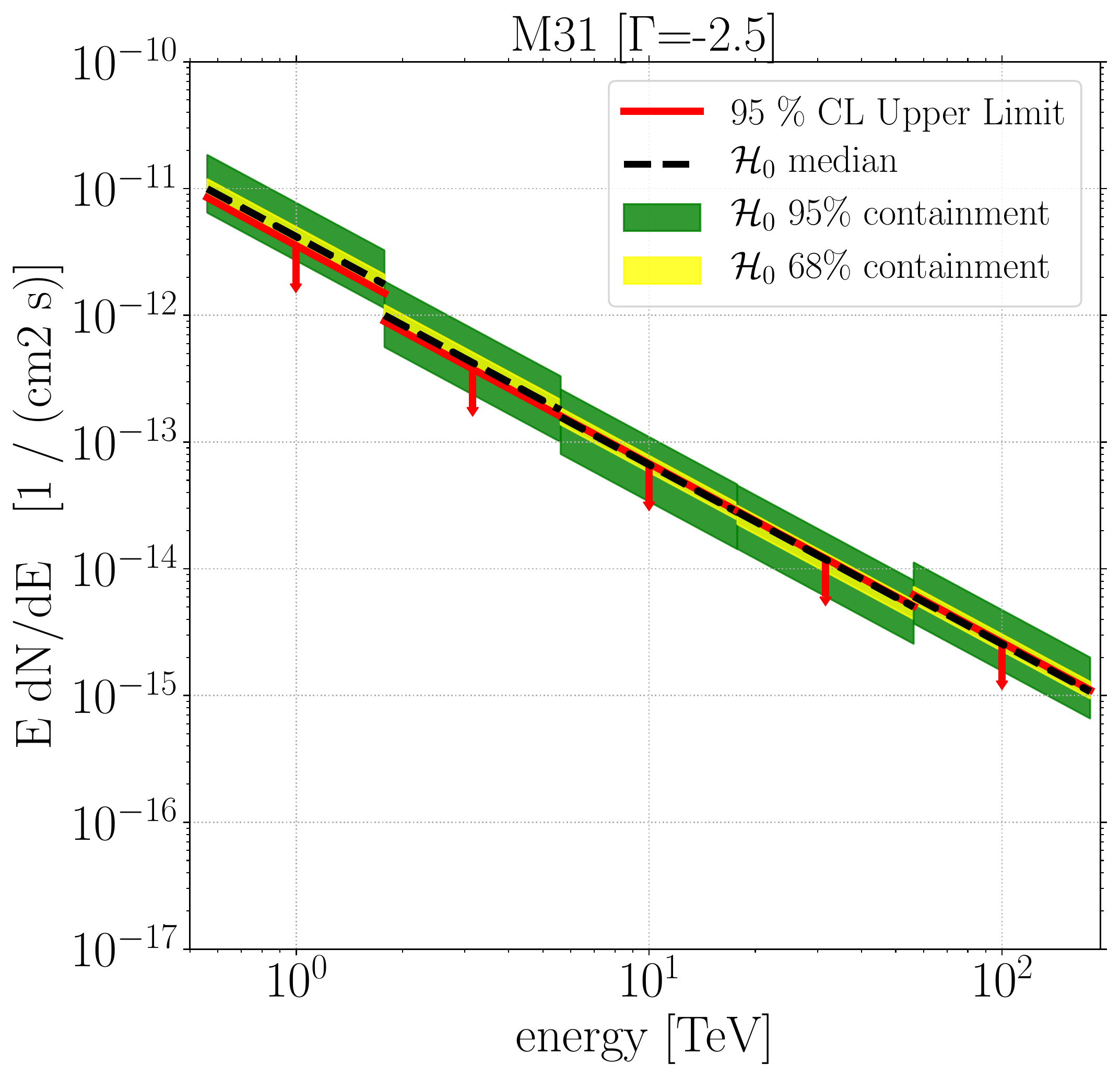}
    \includegraphics[width=.32\textwidth]{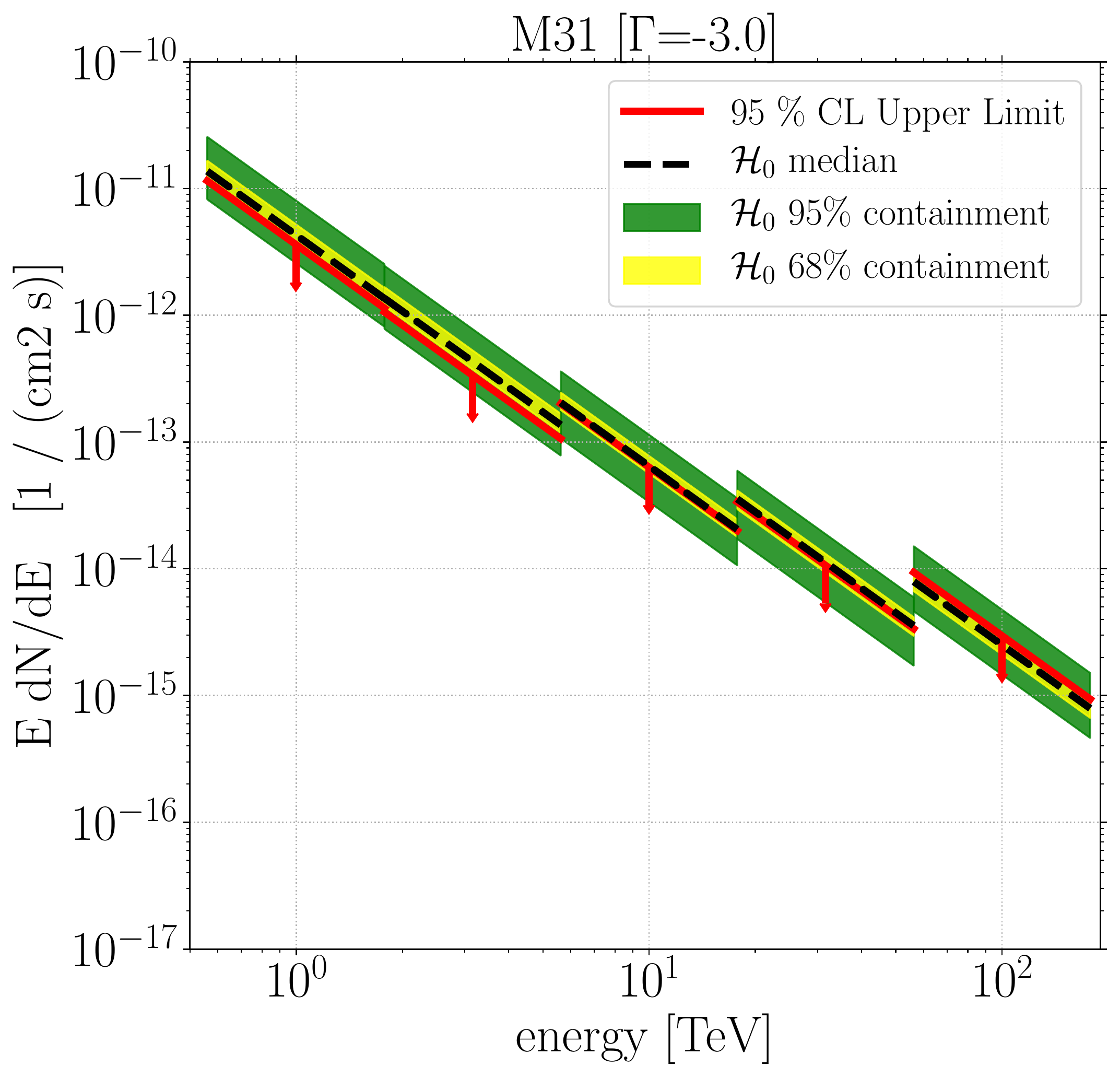}
    \caption{Measured quasi-differential upper limits and expected limits for the M31 disk for three assumed spectral indices}
    \label{fig:quasi_m31_appendix}
  \end{figure}

  \begin{figure}[h!]
    \centering
    \includegraphics[width=.32\textwidth]{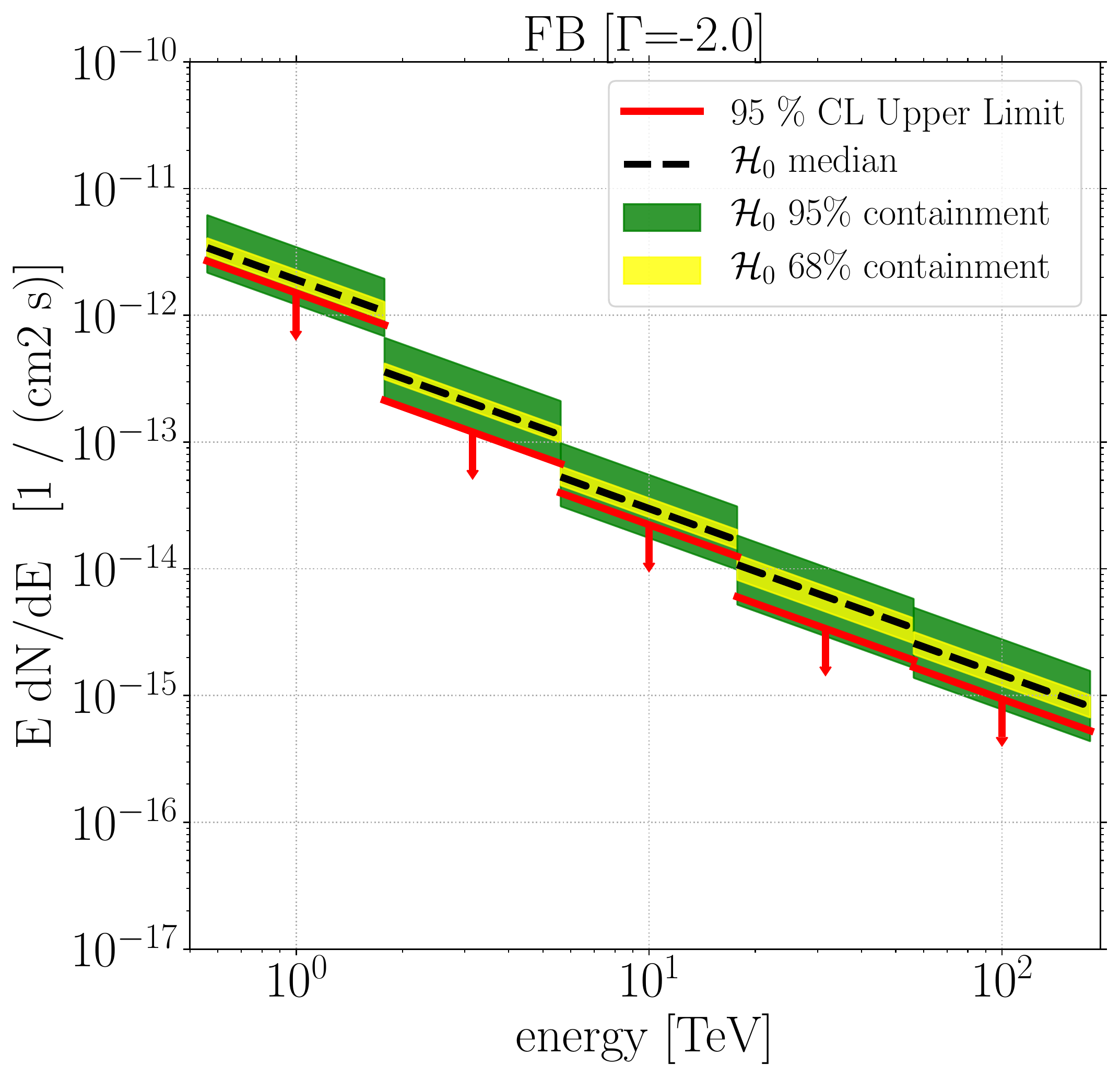}
    \includegraphics[width=.32\textwidth]{quasi_FB_2p5_ednde.pdf}
    \includegraphics[width=.32\textwidth]{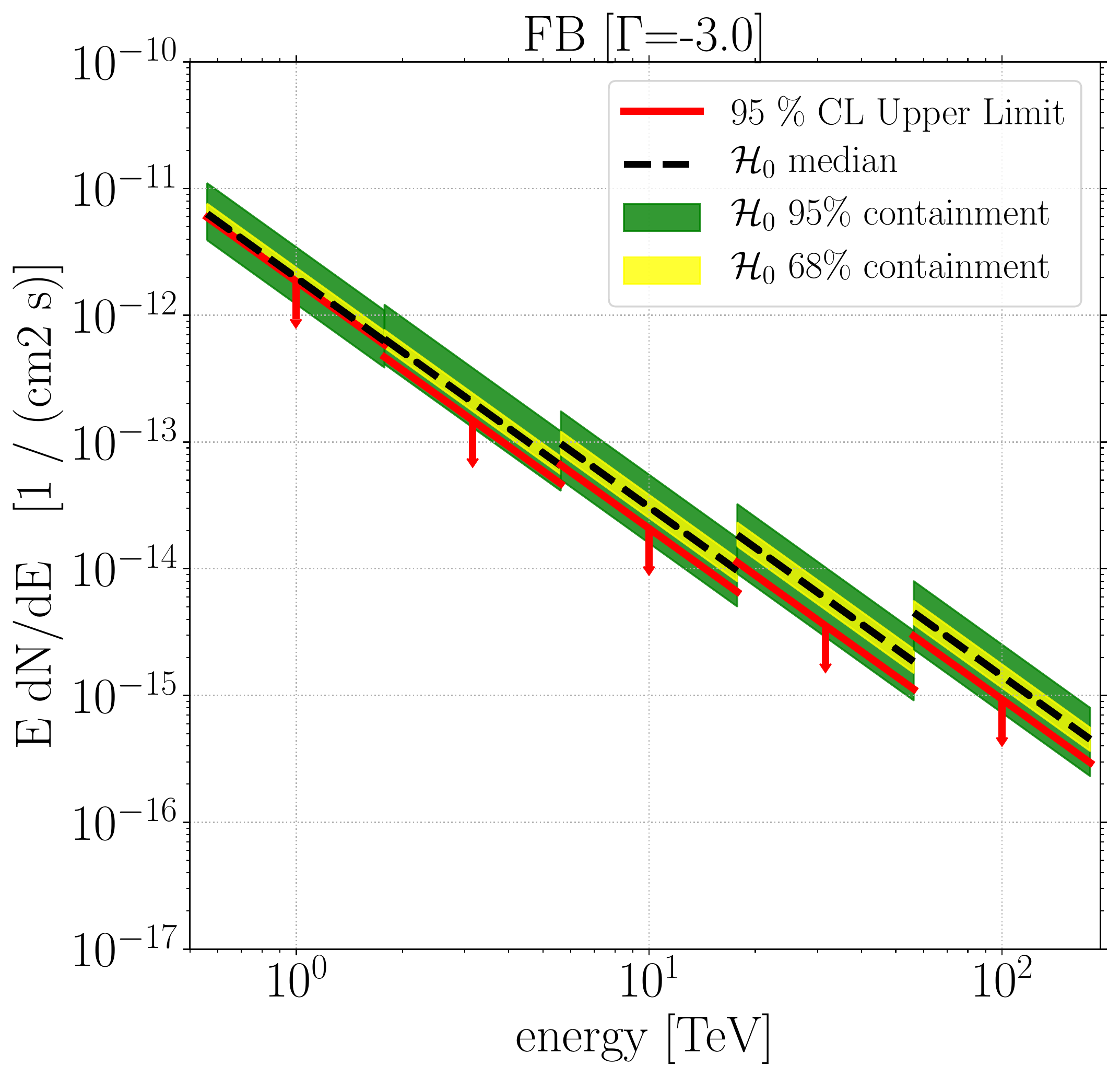}
    \caption{Measured quasi-differential upper limits and expected limits for the M31 FBs for three assumed spectral indices.}
    \label{fig:quasi_bubbles}
  \end{figure}

  \begin{figure}[h!]
    \centering
    \includegraphics[width=.32\textwidth]{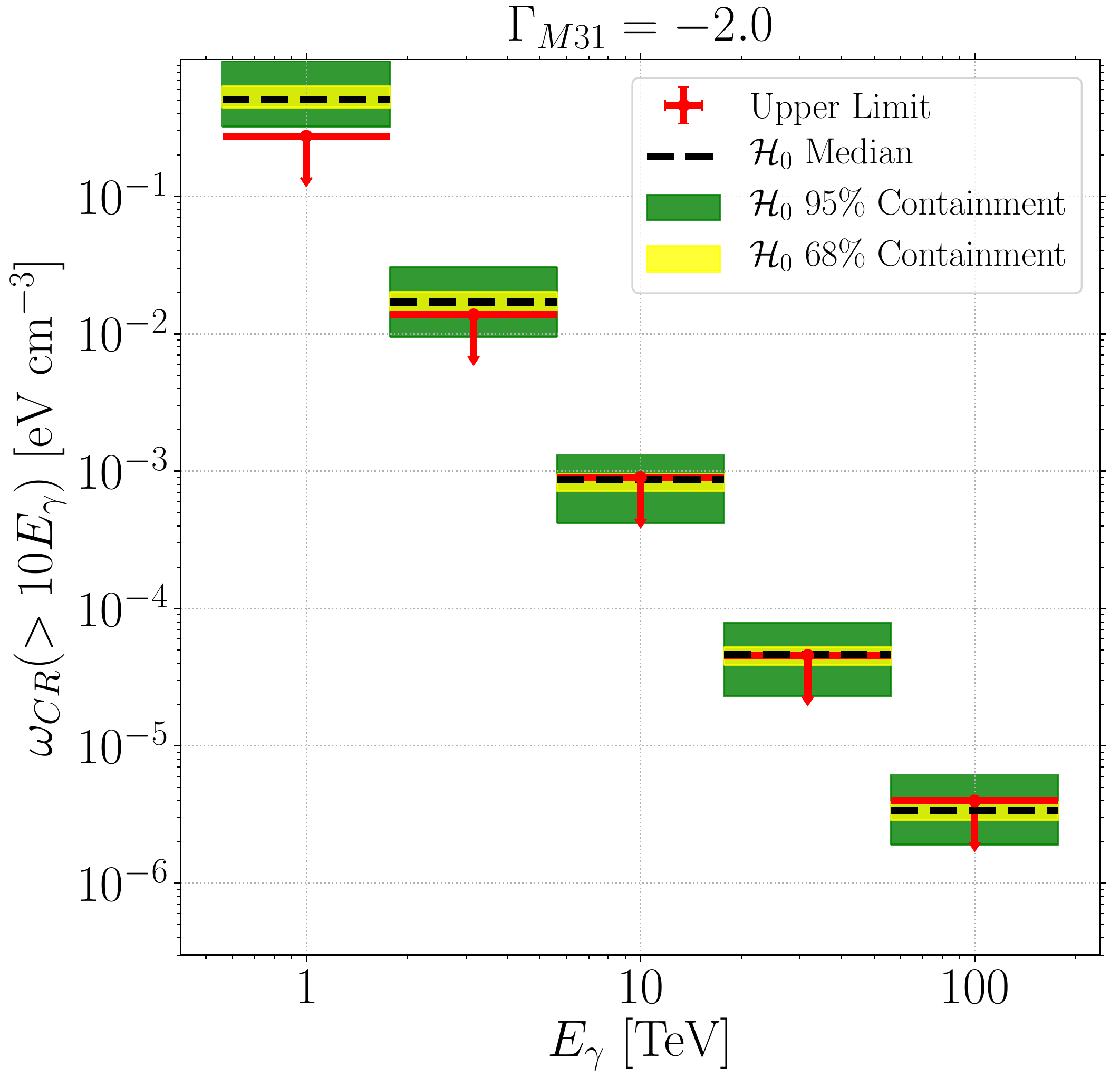}
    \includegraphics[width=.32\textwidth]{cosmic_ray_energy_density_index2p5.pdf}
    \includegraphics[width=.32\textwidth]{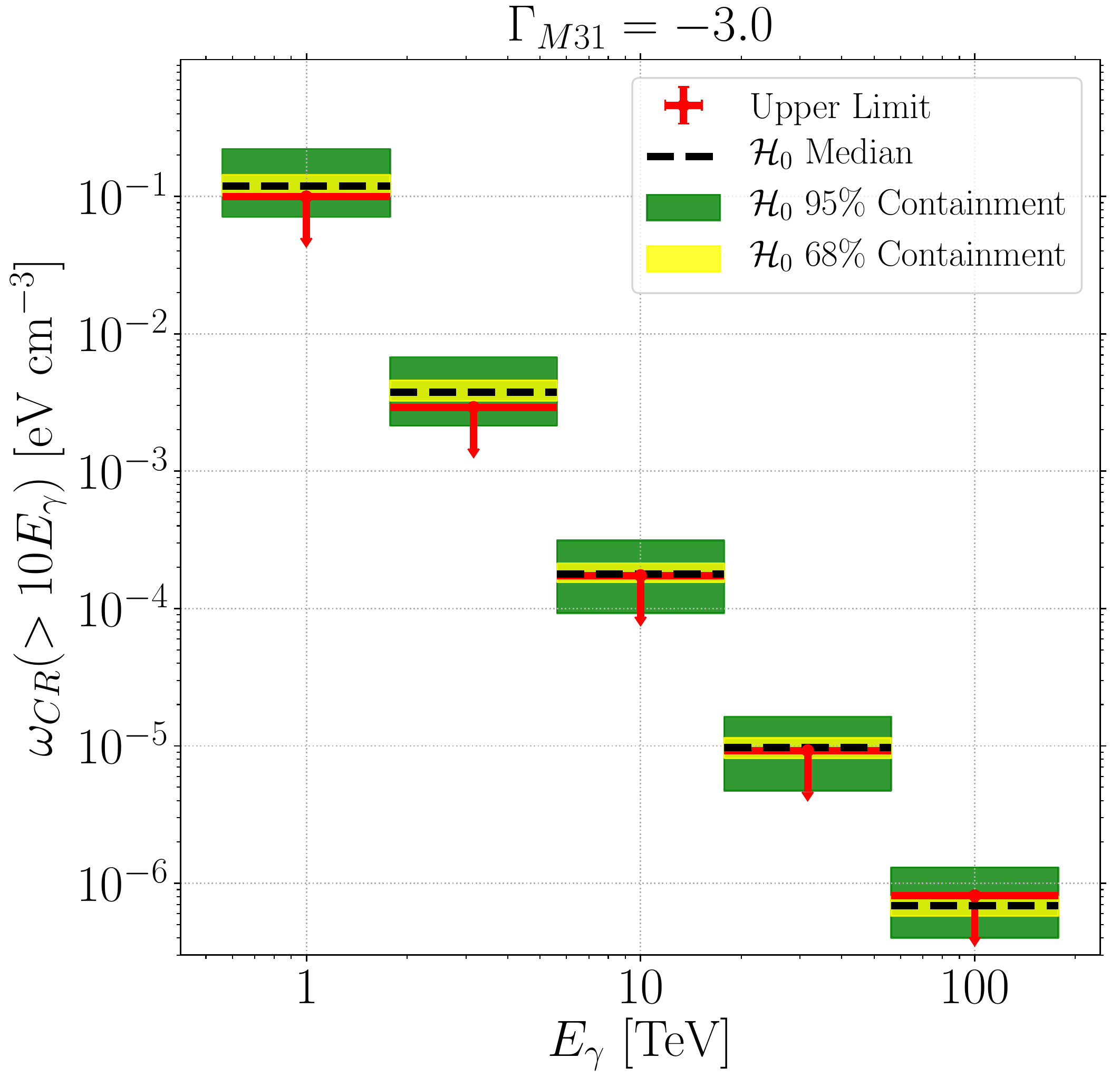}
    \caption{HAWC expected upper limits for cosmic-ray energy density for different spectral indices.}% The Milky way cosmic-ray energy density \citep{Abramowski:2016mir} is shown for comparison.}
    \label{fig:cosmicray_sensitivity_appendix}
  \end{figure}
  
  \begin{sidewaystable}
    \centering
    \begin{tabular*}{\textwidth}{@{\extracolsep{\fill}} c c c c }
    Model    & K                                  & K$_\text{expected}$
    & K$_\text{best fit}$  \\
    Template & [$10^{-14}\text{ cm}^{-2}\text{ s}^{-1}\text{ TeV}^{-1}$] & [$10^{-14}\text{ cm}^{-2}\text{ s}^{-1}\text{ TeV}^{-1}$] & [$10^{-14}\text{ cm}^{-2}\text{ s}^{-1}\text{ TeV}^{-1}$] \\
    \hline \hline
    M31 Disk ($\Gamma = -2.0$) & 8.50 & 9.34 (5.14-16.47, 7.69-10.94) & 3.52 (2.63) \\
    M31 Disk ($\Gamma = -2.5$) & 56.26 & 57.42 (31.27-90.48, 47.77-69.46) & 21.76 (17.60) \\
    M31 Disk ($\Gamma = -3.0$) & 103.59 & 141.91 (88.16-257.09, 119.96-175.46) & 38.96 (32.44) \\
    \hline
    FB ($\Gamma = -2.0$) & 3.11 & 4.99 (2.45-8.57, 3.88-6.31) & 1.06 (0.93) \\
    FB ($\Gamma = -2.5$) & 25.08 & 28.72 (15.15-49.65, 24.29-34.02) & 7.19 (8.24) \\
    FB ($\Gamma = -3.0$) & 40.67 & 65.53 (39.17-126.04, 57.22-80.70) & 14.73 (13.06) \\
    \hline\hline
    Combined ($\Gamma_{M31} = -2.0, \Gamma_{FB} = -2.0$) & 7.80 & 7.49 (3.13-16.35, 6.24-9.09) & 2.84 \\
                                                           & 3.38 & 4.46 (2.17-7.39, 3.48-5.71) & 1.12 \\
    \hline
    Combined ($\Gamma_{M31} = -2.0, \Gamma_{FB} = -2.5$) & 8.46 & 7.76 (3.85-16.02, 6.41-9.41) & 3.38 \\
                                                           & 15.36 & 22.20 (8.53-44.00, 17.03-31.40) & 4.73 \\
    \hline
    Combined ($\Gamma_{M31} = -2.0, \Gamma_{FB} = -3.0$) & 11.67 & 8.70 (4.52-16.27, 6.93-10.16) & 4.60 \\
                                                           & 29.49 & 51.10 (16.09-103.44, 38.80-77.28) & 9.65 \\
    \hline
    Combined ($\Gamma_{M31} = -2.5, \Gamma_{FB} = -2.0$) & 39.00 & 38.07 (11.39-88.72, 29.03-53.68) & 13.34 \\
                                                           & 3.73 & 4.47 (2.36-7.98, 3.67-5.90) & 1.31 \\
    \hline
    Combined ($\Gamma_{M31} = -2.5, \Gamma_{FB} = -2.5$) & 46.96 & 44.23 (22.50-89.59, 36.67-55.52) & 15.98 \\
                                                           & 16.03 & 24.61 (11.43-43.72, 19.57-32.11) & 5.69 \\
    \hline
    Combined ($\Gamma_{M31} = -2.5, \Gamma_{FB} = -3.0$) & 51.07 & 48.77 (19.78-88.90, 38.29-58.56) & 19.56 \\
                                                           & 42.67 & 56.60 (25.44-109.20, 45.62-70.00) & 16.89 \\
    \hline
    Combined ($\Gamma_{M31} = -3.0, \Gamma_{FB} = -2.0$) & 98.67 & 97.25 (36.15-234.42, 76.61-119.26) & 26.26 \\
                                                           & 3.18 & 4.93 (2.57-8.34, 4.07-6.06) & 1.22 \\
    \hline
    Combined ($\Gamma_{M31} = -3.0, \Gamma_{FB} = -2.5$) & 99.92 & 107.32 (46.34-227.66, 80.12-131.09) & 38.04 \\
                                                           & 15.71 & 27.00 (10.29-45.41, 21.57-33.21) & 5.09 \\
    \hline
    Combined ($\Gamma_{M31} = -3.0, \Gamma_{FB} = -3.0$) & 110.85 & 109.35 (58.16-227.98, 91.45-132.16) & 36.28 \\
                                                           & 48.55 & 57.82 (32.36-100.36, 48.41-71.69) & 18.41 \\
    \hline
    \end{tabular*}
    \caption{Normalization values for for 95\% credible interval upper limits, expected limits and the best fit on the emission from the M31 galactic disk and its bubbles. For the combined model results, 95\% credible interval upper limits for M31 and FBs are given seperately, respectively.}
    \label{tab:limits2}
  \end{sidewaystable}  

\end{appendix}

\end{document}